\documentclass[usegraphicx,useAMS,usenatbib]{mn2e}
\usepackage{graphicx}
\usepackage{ulem}
\usepackage{amssymb}
\usepackage{amsmath}
\usepackage{soul}
\usepackage{color}
\usepackage{aas_macros}
\usepackage{subfigure}

\def\fig{Fig.~}
\def\eq{Eq.~}

\def\sims{{\it Box2/hr}}

\def\mfive{M_{500}}
\def\rfive{R_{500}}

\def\msunh{\,h^{-1}\rm{\,M_{\odot}}}

\def\ergs{\rm{\,erg/s}}

\def\cm{\rm{\,cm}}
\def\ks{\rm{\,ks}}
\def\kms{\rm{\,km/s}}

\def\mpch{\,h^{-1}\rm{\,Mpc}}

\def\om{\Omega_{\rm 0M}}

\def\oml{\Omega_{\rm 0\Lambda}}

\def\lbol{\rm{\,L_{\rm bol}}}
\def\lrad{\rm{\,L_{\rm rad}}}
\def\lsxr{\rm{\,L_{\rm SXR}}}
\def\lhxr{\rm{\,L_{\rm HXR}}}

\def\xspec{{\tt XSPEC}}
\def\vapec{{\tt VAPEC}}

\def\wabs{{\tt WABS}}

\def\aap{A\&A}

\def\apjl{ApJ}

\def\apjs{ApJS}

\title[Impact of AGN contamination on ICM X-ray emission]{AGN contamination of galaxy-cluster thermal X-ray emission: predictions for eRosita from cosmological simulations}

\author[V. Biffi, K. Dolag \& A. Merloni]{
  V. Biffi,$^{1,2}$\thanks{E-mail: biffi@oats.inaf.it (VB)}
  K. Dolag$^{3,4}$
  and A. Merloni$^{5}$
  \\
  $^1$ Dipartimento di Fisica dell' Universit\`a di Trieste,
  Sezione di Astronomia, via Tiepolo 11, I-34131 Trieste, Italy\\
  $^2$ INAF, Osservatorio Astronomico di Trieste, via Tiepolo 11,
  I-34131, Trieste, Italy\\
  $^3$ University Observatory Munich, Scheinerstra\ss{}e 1,
  D-81679 Munich, Germany\\
  $^4$ Max-Planck-Institut f\"ur Astrophysik,
  Karl-Schwarzschild-Stra\ss{}e 1, 85748 Garching bei M\"unchen, Germany\\
  $^5$ Max-Planck-Institut f\"ur extraterrestrische Physik,
  Giessenbachstra\ss{}e 1, D-85748 Garching bei M\"unchen, Germany
}
\begin{document}

\pagerange{\pageref{firstpage}--\pageref{lastpage}} \pubyear{...}
\maketitle
\label{firstpage}

\begin{abstract}
  In this study, we present a modelling of the X-ray emission from the
  simulated SMBHs within the cosmological hydrodynamical {\it
    Magneticum Pathfinder Simulation}, in order to study the
  statistical properties of the resulting X-ray Active Galactic Nuclei
  (AGN) population and their expected contribution to the X-ray flux
  from galaxy clusters.  The simulations reproduce the evolution of
  the observed unabsorbed AGN bolometric luminosity functions (LFs) up
  to redshift $z\sim2$, consistently with previous works.
  Furthermore, we study the evolution of the LFs in the soft (SXR) and
  hard (HXR) X-ray bands by means of synthetic X-ray data generated
  with the PHOX simulator, that includes an observationally-motivated
  modelling of an instrinsic absorption component, mimicking the torus
  around the AGN. The reconstructed SXR and HXR AGN LFs present a
  remarkable agreement with observational data up to $z\sim2$ when
  an additional obscuration fraction for Compton-thick AGN is assumed,
  although a discrepancy still exists for the
  SXR LF at $z=2.3$.  With this approach, we also generate full
  eROSITA mock observations to predict the level of contamination due
  to AGN of the intra-cluster medium (ICM) X-ray emission, which can
  affect cluster detection especially at high redshifts.  We find
  that, at $z\sim 1\mbox{--}1.5$, for $20\mbox{--}40\%$ of the
  clusters with $\mfive > 3\times 10^{13}\msunh$, the AGN counts in
  the observed SXR band exceed by more than a factor of 2 the counts
  from the whole~ICM.
\end{abstract}
\begin{keywords}
  methods: numerical -- X-rays: galaxies: clusters -- galaxies: active -- galaxies: quasars: supermassive black holes
\end{keywords}

%
%
\section{Introduction}\label{sec:intro}

The presence of supermassive black holes (SMBHs) is nowadays known to
be very common at the centre of massive galaxies and show properties,
such as total mass, that appear to be correlated to the properties of
their hosts, such as bulge stellar mass or velocity dispersion~\cite[e.g.][]{magorrian1998,ferrarese2000,gebhardt2000,tremaine2002,ferrarese2005,kormendy2013,mcconnell2013}.
The nuclear activity due to
matter accretion onto the central SMBH, however, is observed in only
1--10\% of the galaxies, indicating that the co-evolution between
active galactic nuclei (AGN) and host galaxy is a very complex and
intermittent process.

In the case of brightest cluster galaxies (BCGs) in galaxy clusters,
the importance of AGN activity is even more crucial, as it affects not
only the star formation activity within the
BCG~\cite[][]{benson2003,croton2006} but also the wider
environment around it and the hot gas filling the cluster potential
well, i.e. the intra-cluster medium
(ICM)~\cite[][]{fabian2003,mcnamara2000,mcnamara2005,voit2005}.
ICM thermal and chemical properties in the cores of clusters show the
effects of the central AGN, depending on its growth and
evolution~\cite[e.g.][and references therein]{gitti2012}.  In
fact, if no efficient feedback mechanism was in place at the centre of
galaxy clusters, massive cooling flows were to be expected as a
consequence of the radiative cooling of the
ICM~\cite[see][]{fabian1984,sarazin1986,fabian1994}, whose
temperatures reach $10^7$--$10^8$\,K emitting in the X-rays with
typical luminosities of $10^{43\mbox{--}45}\ergs$. As an observable
signature of this process, we should expect large reservoirs of cool
gas in the centre of clusters, especially in those with a high-density core
and centrally peaked X-ray surface brightness (cool-core clusters),
resulting in considerable star formation rates in the BCG. This
so-called ``cooling flow'' scenario~\cite[][]{fabian1994} is however
not observed and a heating mechanism able to prevent such gas
radiative cooling and star formation in the core must be in
place~\cite[][and references
  therein]{peterson2001,david2001,peterson2006}. It is commonly
accepted that this role is played by AGN feedback (see review articles by~\citealt{gitti2012} and \citealt{fabian2012}, and references therein).

Statistically, studies of the populations of AGN at different
redshifts, for instance through the construction of luminosity
functions (LFs) at various wavelenghts, are particularly useful to
investigate the growth history of SMBHs in the Universe and to infer
constraints on their co-evolution with the host galaxy and the
environment. Several observational campaigns have been dedicated to
this scope~\cite[e.g.][]{maccacaro1983,maccacaro1984,maccacaro1991,boyle1993,boyle1994,page1996,boyle2000,wolf2003,ueda2003,simpson2005,barger2005,lafranca2005,richards2006,bongiorno2007,silverman2008,hasinger2008,croom2009,aird2010,aird2015,assef2011,fiore2012,merloni2014,buchner2015,fotopoulou2016,ranalli2016}.

Given the complexity of the phenomena, however, it is very difficult
to capture the details of the cycle regulating the gas inflow towards
the cluster centre, the accretion of this gas onto the central SMBHs,
the triggering of powerful AGN activity and the consequent release of
energy from the AGN into the surrounding medium, that prevents further
gas accretion and quenches star formation.

The challenge to include the modelling of these phenomena within
  simulations is mainly related to the involved time-scales, which
  differ largely --- e.g. the time-step of the simulation compared to
  the time-scale over which the process of gas accretion onto the BH
  (and therefore the following variation of the AGN luminosity)
  occurs.  The estimation of AGN luminosities from cosmological
  simulations that include SMBHs, and account for the modelling of gas
  accretion onto them, should therefore be regarded mainly from a
  statistical perspective.
The possibility to correctly model a realistic population of AGN that
originates directly from the SMBH population within a cosmological
context is nevertheless the key step to study and understand the SMBH
evolution and their impact on physical properties of galaxies and
clusters.

Several theoretical works, both numerical and semi-analytical, have
also focused on the population of SMBHs and their relation to the
small-scale galactic environment, and on the constructions of
luminosity and mass functions that are crucial to be compared against
observational evidences in order to interpret them as well as to
validate the modelling itself.
In particular, cosmological hydrodynamical simulations have been
performed in the past two decades with the specific purpose of
studying the statistical properties of the simulated SMBHs and AGN
populations, such as LFs, within a cosmological
context~\cite[e.g.][]{dimatteo2008,dimatteo2012,mccarthy2010,mccarthy2011,booth2011,degraf2010,degraf2011,haas2013,H14,sijacki2015,lisa2015,lisa2016,OWLS2017},
including various different implementations of the AGN feedback
model~\cite[e.g.][]{dimatteo2005,springeldimatteo2005,hopkins2006,sijacki2007,booth2009,fabjan2010,barai2014,lisa2015}.

Cosmological hydrodynamical simulations of galaxy clusters have also
shown how the presence of AGN feedback, in addition to stellar one,
can impact on the entropy, metallicity and thermal properties in
cluster cores, by reducing the amount of high density and low
temperature gas in the centre which causes excessive star formation.
This eventually contributes to a diversity of thermal and chemical
properties of the cores that resemble the observed
ones~\cite[e.g.][]{hahn2015,rasia2015,martizzi2016,planelles2017,biffi2017,vogelsberger2017}.

The comparison between numerical results and observational data is
crucial in order to assess the reliability of the models included in
the simulations and to interpret the observational results themselves,
although comparison studies of this kind have been relatively limited
so far.  For a more direct and faithful comparison,
synthetic data can be derived from the simulations.
In a recent work, \cite{OWLS2017} derived mock X-ray AGN
catalogs from the cosmo-OWLS simulation suite, in order to investigate
the demographics of the AGN population in the simulations.  They find
that the unabsorbed X-ray luminosity function accurately reproduces
the observed one over 3 orders of magnitude in X-ray luminosity from
$z=0$ out to $z=3$, as well as the Eddington ratio distribution and
the projected clustering of X-ray AGN.

\looseness=-1 For a realistic population of simulated BHs and AGN, in terms of
comparison to observational evidences, simulations and mock X-ray data
are powerful tools to constrain the possible contribution from AGN
X-ray emission to the ICM emission of the host cluster.
In fact, X-ray observations point out the difficulty in detecting an
active AGN during its outburst phase within the BCG of a galaxy
cluster. This is essentially due to the ambiguity of disentangling the
AGN point source emission from the X-ray peak associated to the
cluster core, especially in centrally-peaked CC clusters and at high
redshift for X-ray telescopes with moderate (more than a few
arcseconds) spatial resolution.  In fact, this will be crucial for
future X-ray survey instruments like eROSITA, for which the PSF in
scanning mode will be large (Half-Energy Width $\sim 26\mbox{--}28"$
and positional accuracy of point sources $\sim 7\mbox{--}15"$) and
will make it very difficult to resolve the central core in CC
clusters.  In such cases, cluster detection can be challenging,
especially at high redshift, since the instrument spatial resolution
will not allow to distinguish the extended diffuse emission of the ICM
and, in case of powerful AGN in cluster BCGs, the X-ray emission will
be likely associated to them rather than to the hosting cluster.
Nevertheless, if the host galaxy of the detected AGN is a member of a
massive cluster, then the X-ray emission from the AGN can be a minor
fraction of the total AGN and ICM emission.  The situation where the
point source is classified as an AGN and the cluster remains
uncatalogued contributes to the rarity of selecting active AGN at the
centre of massive clusters and constitutes an important selection bias
that should be taken into account especially in future surveys. A way
to overcome this problem, by combining X-ray and optical data, is
presented in a recent work by~\cite{green2017}, where they look for
evidences of the presence of a rich cluster around ROSAT X-ray sources
identified as AGN, by searching for overdensities in red-sequence
galaxies.

In such a framework, we aim at investigating the population of SMBHs
in the {\it Magneticum Pathfinder Simulation} (see Section~2) in terms
of X-ray LFs and their evolution with redshift, in comparison to
observational findings. Given a realistic AGN population, we
ultimately intend to use simulations to predict the importance of AGN
contamination to the X-ray luminosity and observed emission from
galaxy clusters. To this scope, we set up a model for the AGN X-ray
emission that accounts for both intrinsic absorption and luminosity-
and redshift-dependent obscuration fractions, and finally construct
synthetic observations of the simulated clusters with AGN. We consider
the instrumental specifications of the up-coming X-ray satellite
eROSITA in order to predict the relative contribution from AGN and ICM
to the observed X-ray fluxes, which will affect the detection of
galaxy clusters out to redshifts 1--1.5.

More specifically, the paper is organized as follows. In Section~3 we
first discuss the theoretical modelling of the AGN bolometric
emission, from which we reconstruct the bolometric LFs, and then apply
observationally-motivated bolometric corrections to derive the AGN
intrinsic luminosities in the soft ([$0.5$--$2$]\,keV) and hard
([$2$--$10$]\,keV) X-ray band (hereafter: SXR and HXR, respectively).
As a step further, we apply an X-ray photon simulator to the
simulations to simultaneously mimic the X-ray emission from both
cluster ICM and AGN sources, following the approach outlined in
Section~4.
In Sections~5 and~6 we present our main results on the reconstructed
SXR and HXR AGN LFs, with and without intrinsic absorption, and on the
relative contribution of AGN in clusters with respect to the X-ray
emission from the diffuse ICM, as expected in particular from eROSITA
observations.
Finally, we draw our conclusions in Section~7.

%
\section{Cosmological hydrodynamical simulations}\label{sims}

In this work we use one of the simulations that are part of the {\it
  Magneticum Pathfinder Simulation}.\footnote{Simulation project
  webpage: {\tt www.magneticum.org}.}  In particular, the simulation
used comprises a comoving volume of $352\mpch$ and a mass resolution
of $m_{\rm DM}=6.9\times10^8\msunh$, for the dark matter (DM)
component, and
$m_{\rm g}=1.4\times10^8\msunh$,
for the gas. We refer to this run as \sims~\cite[see
  also][]{H14}, where {\it hr} denotes the high resolution
of the run (initially resolved with
$2\times 1564^3$ particles).
The cosmology adopted refers to the 7-year WMAP results, that is we
assume $h_0=0.704$, for the scaled Hubble parameter, and $\om=0.272$
and $\oml=0.728$, for the matter and Cosmological constant density
parameters. Also, we set the initial power-spectrum index to $n=0.963$
and normalize it to $\sigma_8=0.809$.

\subsection{The numerical code}\label{sec:code}

\looseness=-1 These simulations have been performed with the parallel
TreePM-SPH code Gadget-3, an extended version of the Gadget-2 code
presented in~\cite{springel2005}.
This includes an entropy-conserving formulation of the
SPH~\cite[][]{springel2002}, a higher order kernel based on the
sixth-order Wendland kernel~\cite[][]{dehnen2012} with 295 neighbors,
a low-viscosity scheme and a treatment for artificial diffusion that
allow for a better treatment of turbulence and gas
mixing~\cite[][]{dolag2005,beck2016}.
The code also accounts for the treatment of a wide range of physical
processes governing the evolution of the baryonic component.
In fact, it accounts for
metal-dependent radiative cooling~\cite[following][]{wiersma2009}, for
the presence of the cosmic microwave background (CMB) and of a
UV/X-ray uniform ionizing background radiation from quasars and
galaxies, as computed by~\cite{haardt2001}.
Star formation and feedback from galactic winds
driven by supernovae (SN) explosions
(with a velocity of $350\kms$) are implemented following
the original formulaiton by~\cite{springel2003},
and a description for black hole (BH) growth and feedback from active
galactic nuclei (AGN) is also incorporated~\cite[][]{fabjan2010}.
The code comprises as well a detailed model for chemical enrichment,
where the production of heavy elements is implemented according to
proper stellar population life-times and yields, for supernovae type
Ia (SNIa) and type II (SNII), and for intermediate- and low-mass stars
in the asymptotic giant branch (AGB)
phase~\cite[see][]{tornatore2003,tornatore2007}.
The metal cooling and enrichment is followed by tracking directly 11
metal species (H, He, C, N, O, Ne, Mg, Si, S, Ca, Fe).
Additionally, we include isotropic (physical) thermal conduction (with $1/20$
of the classical Spitzer value)~\cite[][]{dolag2004} and passive
magnetic fields~\cite[][]{dolag_stasyszyn2009}.

%
\section{Modelling the AGN X-ray emission}\label{sec:agnmodel}

In this section we describe the theoretical estimation of AGN
bolometric, SXR and HXR luminosities from the simulations, and present
results on the bolometric AGN LFs.


\subsection{Theoretical estimation of the AGN luminosity}\label{sec:agnbol}

The bolometric luminosity associated to any AGN source in the
simulation can be calculated starting from the principal BH properties
traced directly by the simulation itself, such as its mass and
(large-scale) accretion rate, and by assuming some efficiency value
($\varepsilon_r$).

In the standard scheme used to estimate the (bolometric) radiated luminosity
from accretion onto a BH, we have:
\begin{equation}\label{eq:lrad_std}
  \lrad = 
  \varepsilon_r \dot{M} c^2,
\end{equation}
where typically $\varepsilon_r = 0.1$~\cite[see e.g.][and references therein]{maio2013}.

\begin{figure*}
\centering
\includegraphics[width=0.33\textwidth,trim=30 0 180 0,clip]{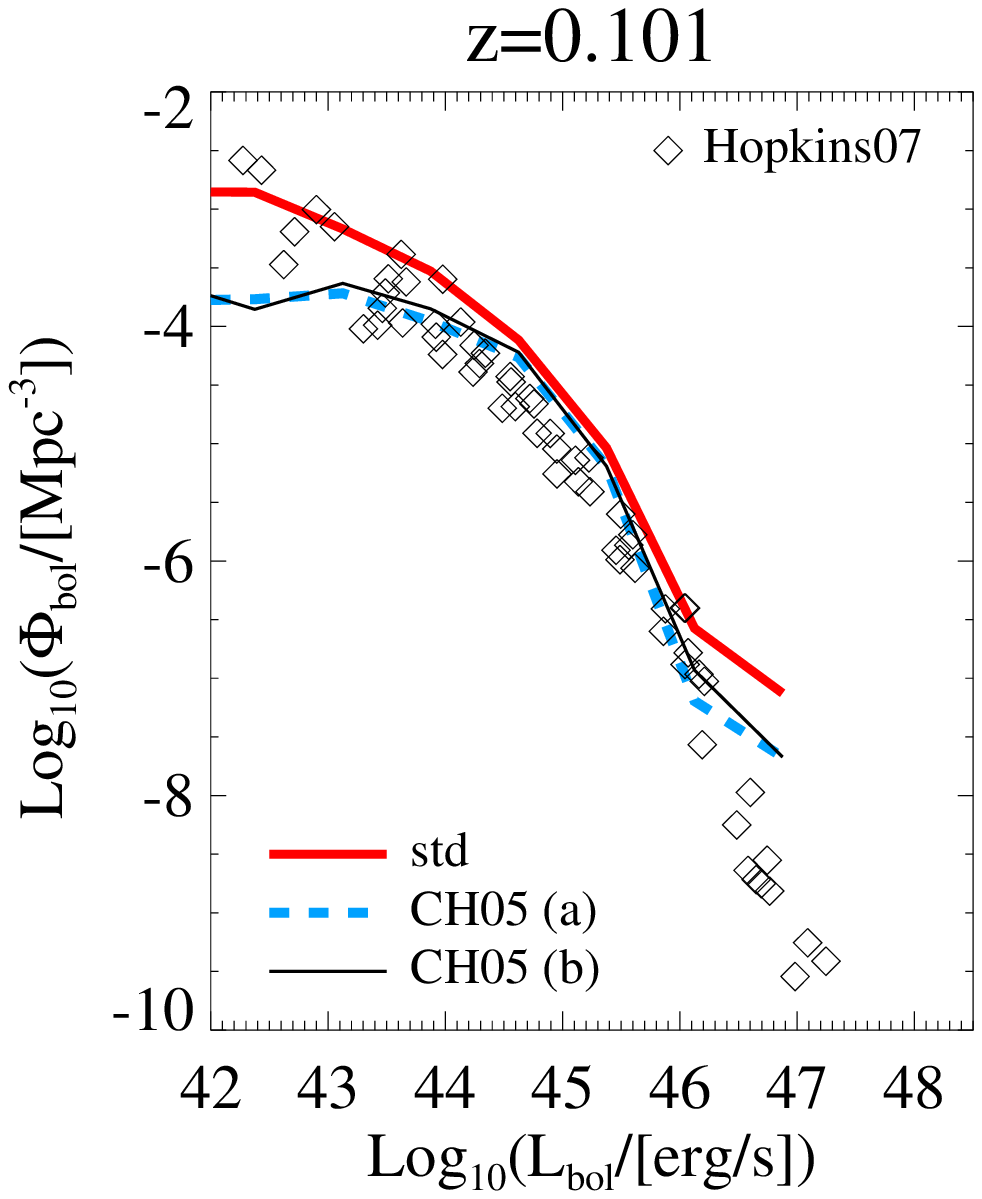}
\includegraphics[width=0.33\textwidth,trim=30 0 180 0,clip]{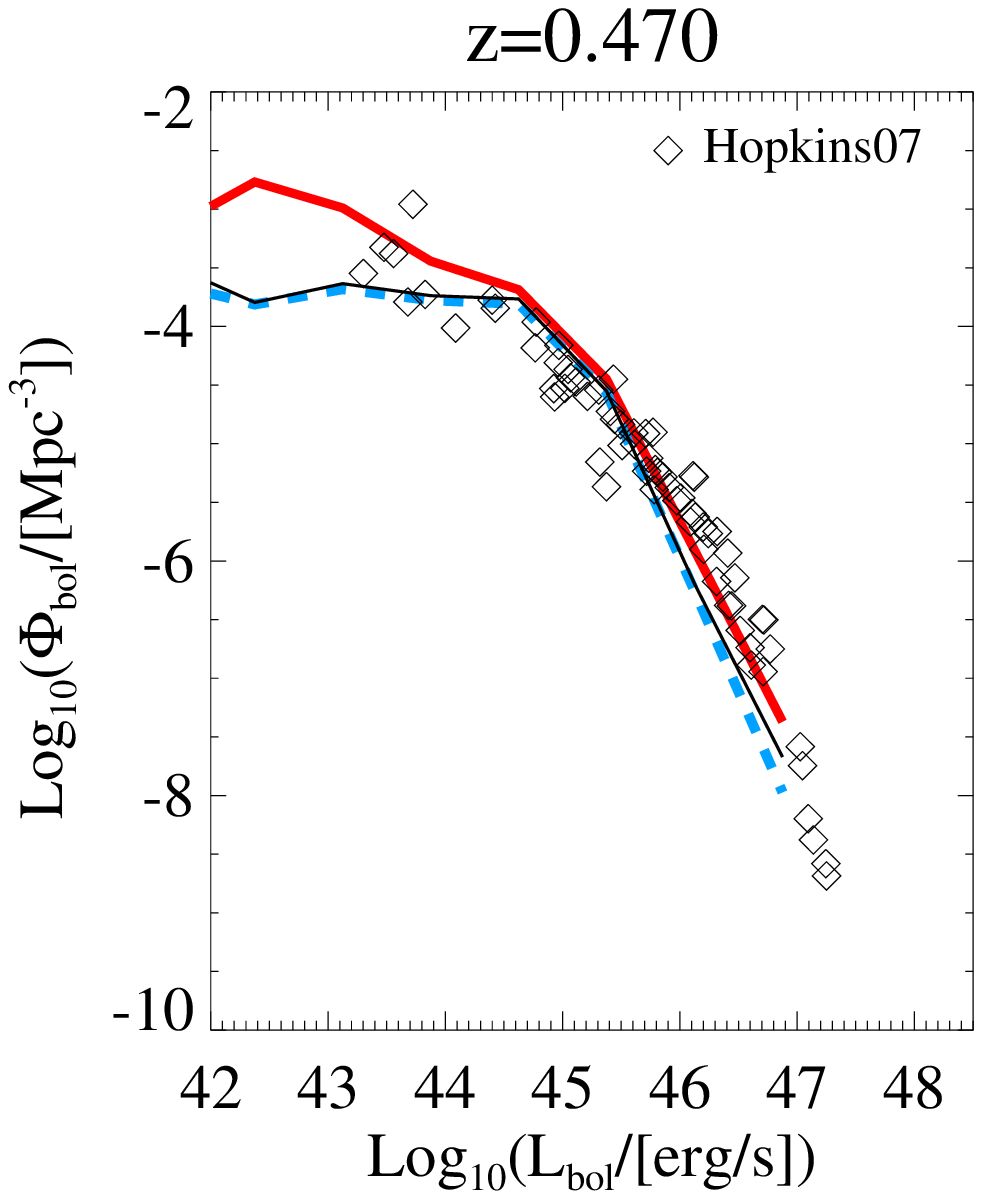}
\includegraphics[width=0.33\textwidth,trim=30 0 180 0,clip]{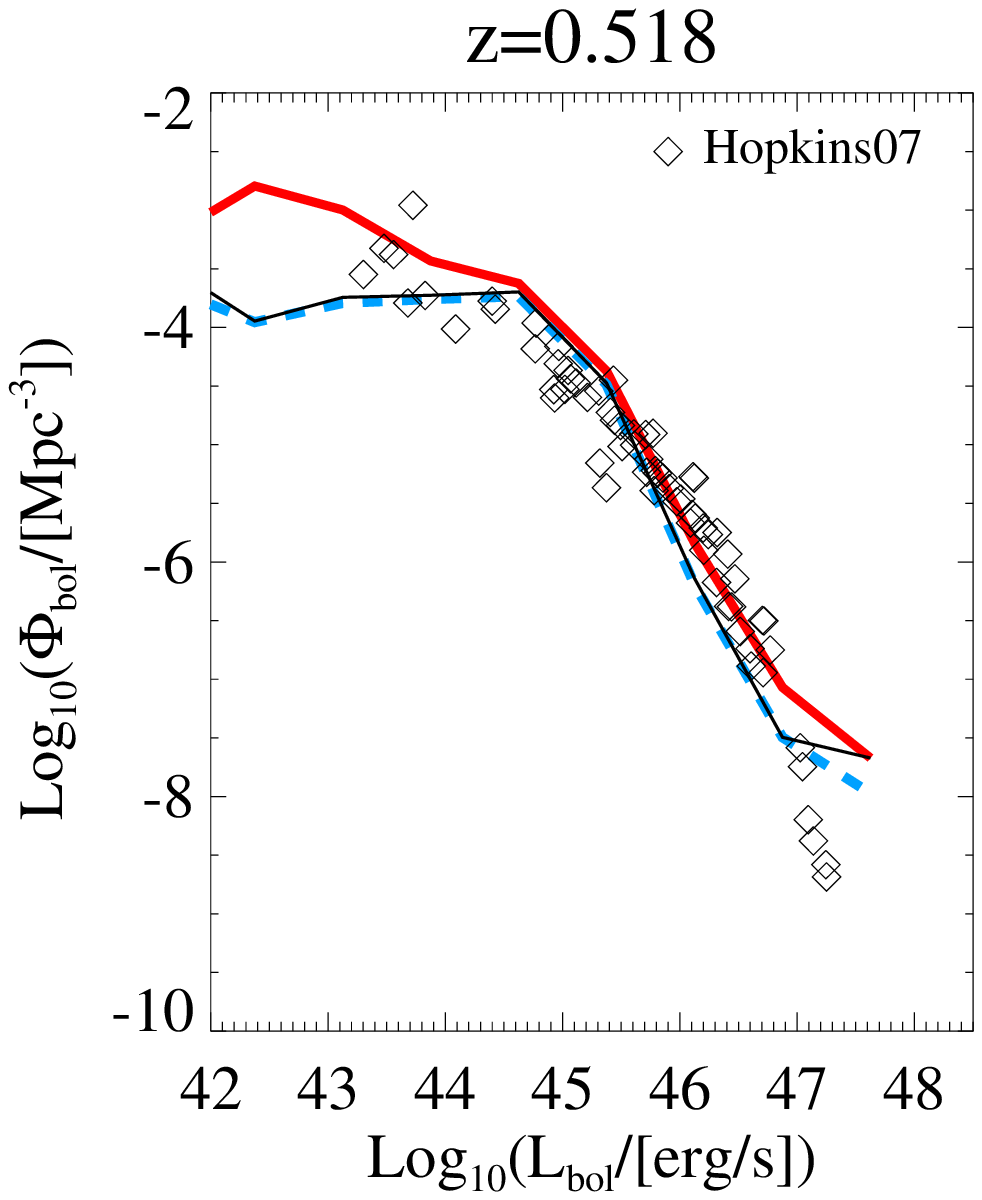}\\
\includegraphics[width=0.33\textwidth,trim=30 0 180 0,clip]{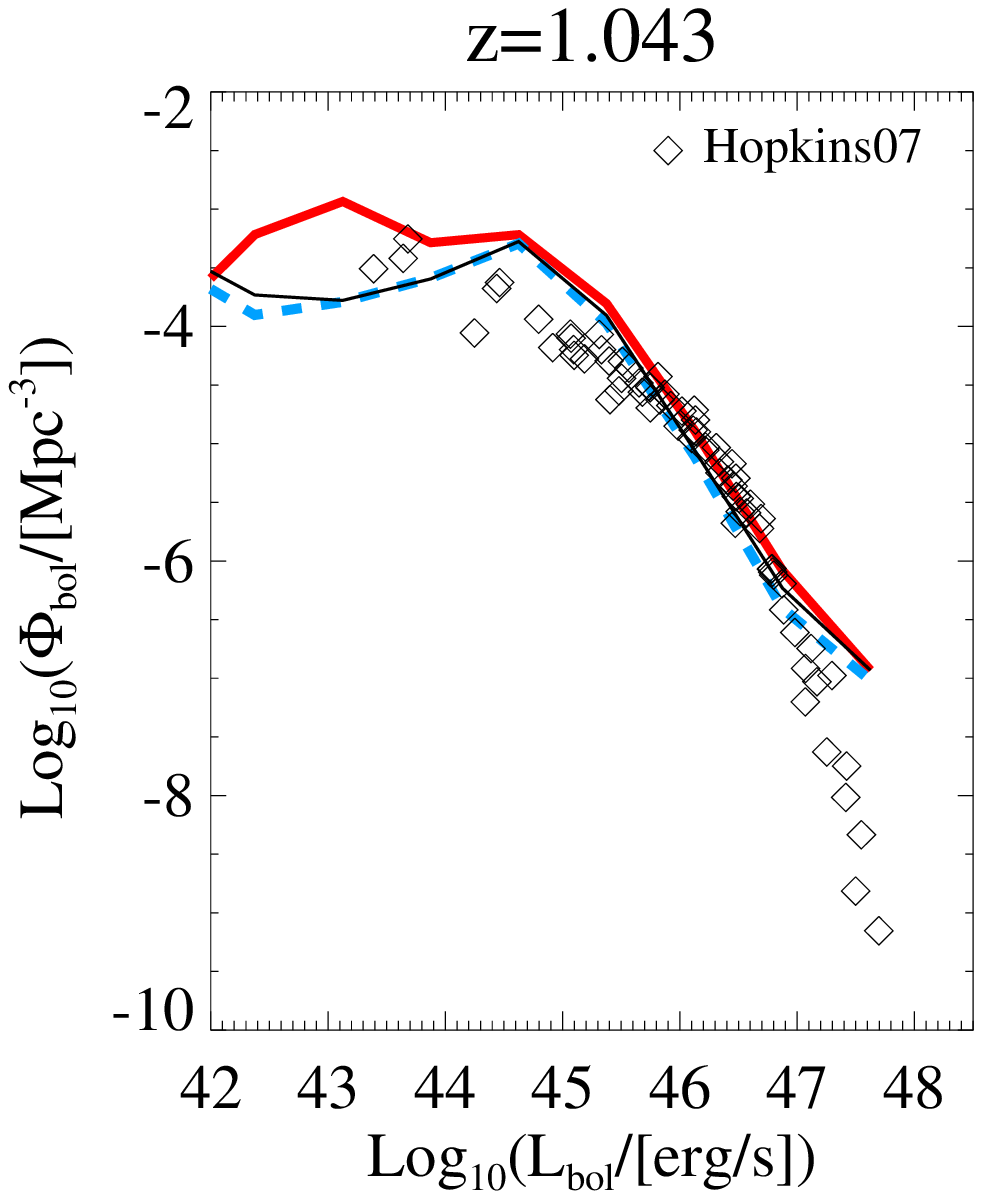}
\includegraphics[width=0.33\textwidth,trim=30 0 180 0,clip]{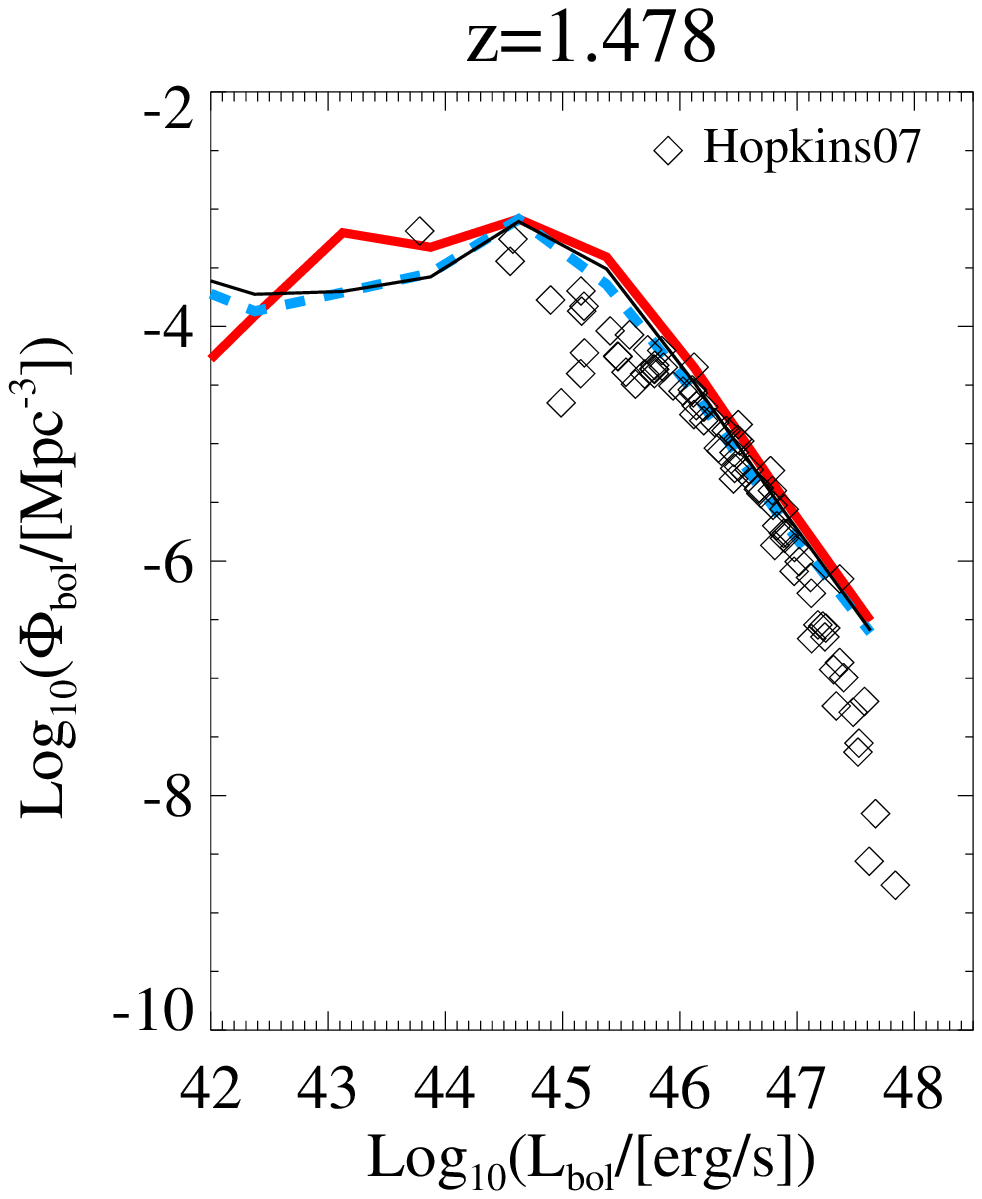}
\includegraphics[width=0.33\textwidth,trim=30 0 180 0,clip]{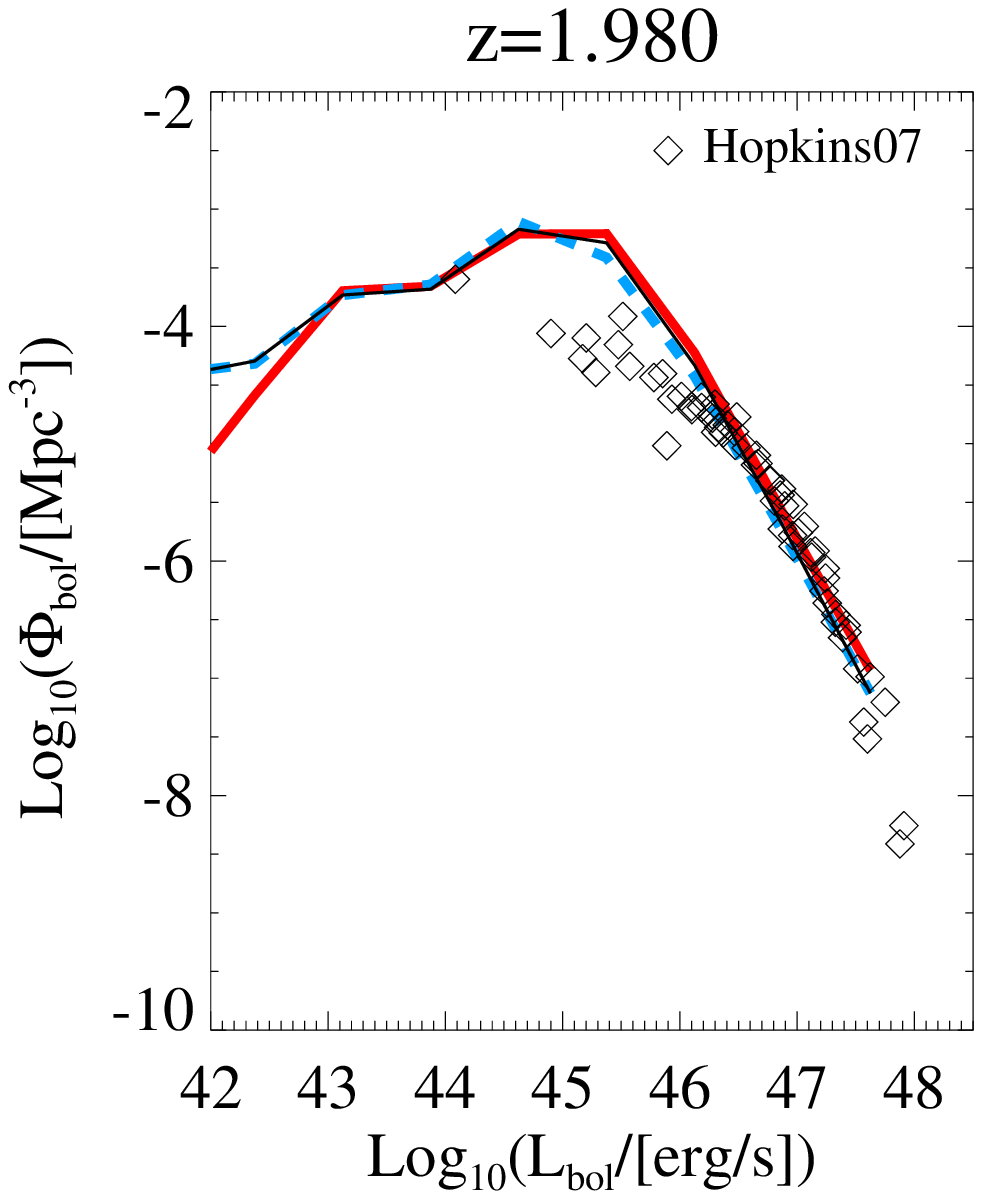}
\caption{Bolometric AGN LFs computed for the \sims\ at redshift $z\sim
  0.1, 0.47, 0.52, 1.04, 1.48, 1.98$ (from top-left to
  bottom-right). Different lines refer to different estimates of the
  AGN bolometric luminosity: the thick solid red line --- ``std'' ---
  corresponds to \eq\eqref{eq:lrad_std}, the dashed blue line
  --- ``CH05~(a)'' --- to \eq\eqref{eq:chur05.2} with a transition
  value for the ${\rm BHAR}$ of 0.1 and the thin solid black line ---
  ``CH05~(b)'' --- to \eq\eqref{eq:chur05.2_005} with a transition
  value for the ${\rm BHAR}$ of 0.05. For comparison we also report
observational data by~\protect\cite{hopkins07} (black
diamonds).\label{fig:bol_xlf}}
\end{figure*}

Alternatively, one can choose a more detailed modelling that takes into account
the specific accretion phase which the AGN source is undergoing, determined according
to its BH accretion rate (BHAR). Namely, the BHAR
\begin{equation}
{\rm BHAR} = \frac{\dot{M}}{\dot{M}_{\rm Edd}}
\end{equation}
is used to distinguish different radiation regimes, following Fig.~1
of~\cite{churazov2005}.  Specifically, the radiative power (i.e.\ AGN
luminosity) is calculated differently for radiatively efficient and
inefficient AGN:\begin{subequations}
\begin{equation}
\frac{\lrad}{L_{\rm Edd}} = \left\{
\begin{aligned}
& \varepsilon _r \left(\!10\frac{\dot{M}}{\dot{M}_{\rm Edd}}\!\right),&
\quad&{\rm if~BHAR}>0.1
\\
& \varepsilon _r \left(\!10\frac{\dot{M}}{\dot{M}_{\rm Edd}}\!\right)^2,&
 &{\rm if~BHAR}<0.1\,
\end{aligned}
\right.
\label{eq:chur05.2}
\end{equation}
corresponding to the radiatively efficient (quasar) (${\rm BHAR}>0.1$) and inefficient (${\rm BHAR}<0.1$) regime, respectively.
We recall that the Eddington quantities read:
\begin{equation}
 L_{\rm Edd} = \frac{4\pi G m_p c}{\sigma_t}M_{\rm BH}
\end{equation}
and
\begin{equation}
 \dot{M}_{\rm Edd} = \frac{L_{\rm Edd}}{\varepsilon_r c^2} = \frac{4\pi G m_p}{\varepsilon_r \sigma_t c}M_{\rm BH}\, .
\end{equation}
Provided the values of $M_{\rm BH}$ and $\dot{M}$ from the simulation,
we only need to assume a value for the efficiency $\varepsilon _r$.
Also in this second approach, we use $\varepsilon _r = 0.1$.

For any BH particle in the simulation, we can therefore compute
its (bolometric) luminosity $\lbol$ and derive
luminosity functions (LFs), in order to explore the properties of the
AGN population.
Here and in the following, we refer to the bolometric luminosity $\lbol$
as the theoretical estimation of
the AGN luminosity, calculated either adopting the standard estimation
or the BHAR-dependent scheme.

The AGN LFs obtained from our simulation are shown in
\fig\ref{fig:bol_xlf}, for different redshifts between $\sim 0.1$ and
$\sim2$.
The three curves reported in the Figure are calculated for the AGN
sources in our simulations considering the standard bolometric
luminosity estimate \eq\eqref{eq:lrad_std} (``std'') or the scheme
by~\cite{churazov2005}. For the latter, we consider either the case
expressed in \eq\eqref{eq:chur05.2} (``CH05 (a)'') or the following
modified version, where we adopt a transition value between radio and
quasar mode of ${\rm BHAR}=0.05$~\cite[see also][]{merloni2008}
(``CH05 (b)''):
\begin{equation}
\frac{\lrad}{L_{\rm Edd}} = \left\{
\begin{aligned}
& \varepsilon _r \left(\!10\frac{\dot{M}}{\dot{M}_{\rm Edd}}\!\right),&
\quad &{\rm if~BHAR}>0.05
\\
& \varepsilon _r \left(\!14.1\frac{\dot{M}}{\dot{M}_{\rm Edd}}\!\right)^2,&
&{\rm if~BHAR}<0.05\,,
\end{aligned}
\right.
\label{eq:chur05.2_005}
\end{equation}\end{subequations}
where we still assume $\varepsilon _r =0.1$.
Being still a standard reference in the field,
we compare the LFs from the {\it Magneticum Pathfinder Simulation}
against observational data by~\cite{hopkins07}, in
which bolometric luminosities are obtained by correcting observed
luminosities in different energy bands.%

From the panels in \fig\ref{fig:bol_xlf} we note that there is an
overall reasonable match in the redshift range $0.47 < z < 2$,
especially in the intermediate-high luminosity end.  At redshift
higher than $1$ we note a tendency to overestimate the observed number
of lower-luminosity clusters.
Similar results were also obtained in~\cite{H14} for the same set of
simulations, where the authors show that a discrepancy w.r.t.\ data rather appears at higher redshifts, $z \gtrsim 2$.

From the comparison in \fig\ref{fig:bol_xlf}, we conclude that in the
luminosity and redshift range where the simulation results are in
reasonable agreement with the observational data, the ``std'' and both
``CH05~(a),(b)'' models do not differ significantly.  At low redshift,
$z\lesssim 1$, the low-luminosity end of the observed LFs is better
reproduced by the ``std'' estimate.

Given these results, and for simplicity, we adopt in the following the
standard estimation given by \eq\eqref{eq:lrad_std} (thick solid red
curves).
%

\subsection{Bolometric corrections: $\lsxr$ and $\lhxr$}\label{sec:agnsxrhxr}

The estimation of $\lbol$ is nevertheless not enough to constrain the
AGN X-ray emission model nor to fully compare the properties of the
simulated AGN population against observed datasets and LFs.  Therefore, it
is particularly useful to derive X-ray luminosities in specific energy
bands, namely the SXR and HXR bands.

To this purpose, we convert the bolometric luminosities into SXR and
HXR luminosities assuming the bolometric corrections proposed
by~\cite{marconi2004}, which can be approximated by the following
third-degree polynomial fits
\begin{align}
\log(\lhxr/\lbol)&=-1.54\!-\!0.24\mathcal{L}\!-\!0.012\mathcal{L}^2 \!+\!0.0015\mathcal{L}^3 \label{eq:lsxr}\\
\log(\lsxr/\lbol)&=-1.65\!-\!0.22\mathcal{L}\!-\!0.012\mathcal{L}^2 \!+\!0.0015\mathcal{L}^3 \label{eq:lhxr}
\end{align}
with $\mathcal{L}=\log(\lbol/L_{\odot})-12$, and derived for the
luminosity range $8.5 < \log(\lbol/L_{\odot}) < 13$~\cite[see
  Fig.~3(b) in][]{marconi2004}.
The reason for this choice is that the corrections are derived
from template spectra, rather than from the average observed
AGN spectral energy distribution (SED), with the goal of obtaining an
estimate of the {\it intrinsic} AGN luminosity. This, in principle, is
closer to the theoretical estimates of $\lbol$ derived from
simulations.
In particular, the spectral model used by~\cite{marconi2004} consists,
in the X-ray band beyond 1\,keV, of a single power law, with a typical
photon index of $\sim 1.9$ and an exponential cut-off at $\sim
500$\,keV, plus a reflection component.
The template spectra, as well as the obtained corrections, are
redshift independent.

Here and in the following, the SXR and HXR bands are considered to be
rest-frame energies, unless otherwise explicitly stated (i.e.\ for the
mock results in Secs.~\ref{sec:res-mocks}--\ref{sec:res-mocks-mass}).

%
%
\section{Synthetic X-ray emission}\label{sec:mocks}

In this section we describe the modelling of the synthetic X-ray ICM
and AGN emission form the simulations, with special attention to the
AGN intrinsic absorption.

\subsection{PHOX X-Ray photon simulator}\label{sec:phox}

In the present work we make use of the PHOX code~\cite[][for further
  details]{biffi2012,biffi2013} in order to generate X-ray synthetic
observations from the ICM and AGN sources in the simulation.
The PHOX X-ray photon simulator consists of three separate modules:
\begin{itemize}
\item
in {\scshape unit\;1} the ideal photon emission in the X-ray band is computed for
every emitting source by assuming a model spectrum, which is sampled
statistically with a discrete number of photons and stored in a data
cube that matches the simulation cube itself;
\item
in {\scshape unit\;2} a projection is applied, photon energies are
Doppler-shifted according to the line-of-sight motion of the original
emitting source and a spatial selection can additionally be
considered;
\item
lastly in {\scshape unit\;3} realistic observing time and detector area are
considered, and the ideal photon list is eventually convolved with the
specific instrumental response of a chosen X-ray telescope.
\end{itemize}

This code has been applied to simulations of galaxy clusters to study
the X-ray properties of the hot diffuse
ICM~\cite[][]{biffi2012,biffi2013,biffi2014,biffi2015,cui2016}, by
modelling the X-ray emission (bremsstrahlung continuum and metal
emission lines) of the hot gas in the simulated clusters.

Nevertheless, the very general approach adopted allows to treat as
well different X-ray sources traced by the simulations, provided that
the corresponding emission model is constrained from the simulation
and included in {\scshape unit\;1}.
Given the modular design of PHOX, the projection and convolution
phases ({\scshape unit\;2} and {\scshape unit\;3}) can be then applied to the photon cube
independently of the different nature of the emitting source.

In this study, we present and employ an extension of PHOX that
generates the sythetic X-ray emission from AGN-like sources included
in the simulations.

\subsection{ICM X-ray emission}\label{sec:phox-icm}

The modelling of the synthetic X-ray emission from the ICM in siulated
clusters is done by computing the emitted X-ray spectrum for every
hot-phase gas element in the simulations, depending on its intrinsic
thermal and chemical properties (namely density, temperature, global
metallicity or singular element abundances). In particular, we assume
for every gas element a single-temperature \vapec\footnote{See {\tt
    http://heasarc.gsfc.nasa.gov/xanadu/xspec/manual/XS
    modelApec.html}.}  thermal emission model with emission lines due
to heavy elements~\cite[][]{apec2001}, implemented in
\xspec\footnote{See {\tt
    http://heasarc.gsfc.nasa.gov/xanadu/xspec/}.}~\cite[][]{xspec1996}.

For further details on the modelling of the ICM X-ray emission, we
refer the interested reader to the descriptions provided
in~\cite{biffi2012,biffi2013}.

\subsection{AGN X-ray emission: spectrum parameters}\label{sec:phox-agn}
The modelling of the AGN X-ray emission in the PHOX code follows a
very similar approach to the one implemented for the ICM emission.
Given the BH properties calculated in the simulation (as described in
Sections~\ref{sec:agnbol} and~\ref{sec:agnsxrhxr}), for each AGN
source element we can constrain the particular shape of the emission
model spectrum and generate the associated (ideal) list of emitted
X-ray photons.\footnote{Simulation data from the {\it
    Magneticum Pathfinder Simulation} and the associated ICM and AGN
  synthetic X-ray data have been made available through
  the Cosmological Web Portal presented in~\cite{ragagnin2017} ({\tt
    http://c2papcosmosim.srv.lrz.de/}).}

The main difference in treating BH sources, instead of gas elements,
resides in the different X-ray emission model expected.  Instead of
modelling the emission like a thermal bremsstrahlung continuum with
metal emission lines, such as for the ICM, we need to assume a
different spectral model.
X-ray observations show evidence that all AGN sources share an
intrinsic power law spectrum of the form $ E^{-\alpha}$, with a photon
index $\alpha$ \cite[e.g.][]{Zdziarski2000}
that varies in the range 1.4--2.8 with a Gaussian distribution peaked around $\sim 1.9$--$2$.
Thus, we assume a single, redshifted power law:
\begin{equation}
A(E) = [K(1+z)] [E(1+z)]^{-\alpha} \left(\frac{1}{1+z}\right),
\label{eq:agnspec}
\end{equation}
where $K$ is the normalization, $\alpha$ is the photon index (spectrum
slope) and $z$ is the redshift of the
source.\footnote{\eq\eqref{eq:agnspec} adopts the notation of the
  redshifted power-law model defined within the \xspec\ package.}
Integrating the spectrum in \eq\eqref{eq:agnspec} between two
observed energies, $E_1$ and $E_2$, one can obtain the observed flux
(and then luminosity) of the source.

In order to calculate the fiducial model for every BH source
(represented by particles in the Lagrangian-based simulations we use
here) in the simulation output, it is required to estimate the
spectrum normalization and slope parameters, $K$ and $\alpha$.  For
each BH, these can be directly constrained from its the global
properties, as sketched in the following.
\begin{itemize}
\item For any BH particle, we compute $\lbol$ from the simulation data
  and convert it to (rest-frame) $\lsxr$ and $\lhxr$ through the
  bolometric corrections~\eqref{eq:lsxr} and~\eqref{eq:lhxr}
  (Secs.~\ref{sec:agnbol} and~\ref{sec:agnsxrhxr}).
\item We assume a spectrum for an AGN-like source as in
  \eq\eqref{eq:agnspec}, so that the luminosity for a given
  (observed) $(E^1,E^2)$ energy band is given by:
\begin{equation}\label{eq:lum}
L_{\left(E^1,E^2\right)} = F \int_{E^1}^{E^2} [K(1+z)] [E(1+z)]^{-\alpha}\!\left(\frac{1}{1+z}\right)\! E {\rm d}E,
\end{equation}
where F is the rescaling factor to convert the flux (resulting from the integration) into the luminosity.
\item From \eq\eqref{eq:lum}, the luminosities in the (restframe) SXR and HXR energy bands are given by:
\begin{equation}
\left\{\!\!
\begin{aligned}
\lsxr &= F \!\int_{0.5/(1+z)}^{2/(1+z)} [K(1+z)] [E(1+z)]^{-\alpha}\!\left(\frac{1}{1+z}\right)\!E {\rm d}E\\
\lhxr &= F \!\int_{2/(1+z)}^{10/(1+z)} [K(1+z)] [E(1+z)]^{-\alpha}\!\left(\frac{1}{1+z}\right)\!E {\rm d}E
\end{aligned}
\right.\!\label{eq:kalpha_1}
\end{equation}
where $\lsxr$ and $\lhxr$ are the values estimated from the simulation.
In order to mimic the observed scatter in these relations, we also add
to both $\lsxr$ and $\lhxr$ a Gaussian scatter with $\sigma=0.1$, in
logaritmic scale.
\item
Solving the system of equations~\eqref{eq:kalpha_1}, we can
constrain the specific values for $K$ and $\alpha$.
\end{itemize}

\begin{figure}
\centering
\includegraphics[width=0.47\textwidth,trim=0 0 0 15,clip]{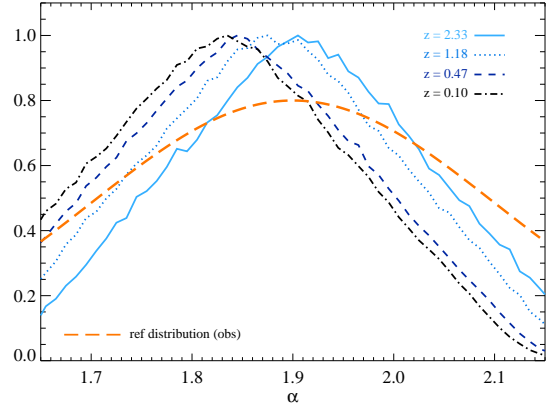}
\caption{Distribution of the slopes computed for the emission spectra
  of the AGN in the simulations used in PHOX~{\scshape unit\;1}, for
  various redshifts between $0.1$ and $2.3$, as in the legend. For comparison, we also
  report the typical distribution expected from observations (orange dashed line).}
\label{fig:alphas}
\end{figure}

In Fig.~\ref{fig:alphas} we show the reconstructed distributions of
slopes found by our approach from the AGN in the simulation box
analysed with PHOX.  In the figure, we show the distribution for 4
example simulation snapshots between $z\sim0.10$ and $z\sim2.3$.  The
result seems in broad agreement with the typical expectations from the
observed distribution (which we consider centered on $\alpha=1.9$ ---
long-dashed orange line).
We note nevertheless a mild shift of the distribution peak towards
lower $\alpha$ values from $z\sim2$ to $z\sim0$. This is likely
related to the redshift evolution of the LFs, which are used together
with bolometric corrections to constrain the~slope.

By reconstructing the normalization and slope, $K$ and $\alpha$,
for every BH we are able to compute the
corresponding AGN spectrum by adopting the {\tt zpowerlw} model
embedded in the \xspec\ package. This can be assumed to well represent
the intrinsic power-law emission of the source in the energy range
interesting for this work ($<50\,$keV).

\subsubsection{Absorption}\label{sec:phox-obsc}

As suggested by many observational works, AGN sources also show
evidences for the presence of obscuring material (i.e. the torus) in
the vicinity of the central BH, which causes a partial absorption of the
emitted radiation.  Therefore, any modelling using a pure power law is
too simplistic to mimic the observed properties of AGN and, even in
the simplest case, it is required to account for an intrinsic
absorption component.

Indeed, the bolometric AGN luminosities from our simulations have been
compared in Section~\ref{sec:agnbol} to the observational results
by~\cite{hopkins07}, in which the authors applied corrections to
account for absorption and convert the observed luminosity into the
intrinsic, bolometric one.
Yet, in order to compare the SXR and HXR LFs of
the AGN from the simulation to observed data, the correction for
--- or modelling of --- obscuration is particularly important.

Several observational studies have focused on
the characterization of the absorption depending on the source
luminosity, and redshift~\cite[see, e.g., studies
  by][]{hasinger2008,merloni2014,buchner2014,buchner2015,fotopoulou2016,ranalli2016}.  The
correction of instrinsic luminosities predicted by numerical studies
by taking into account such observationally-motivated obscuration
fractions has been for instance employed by~\cite{H14}, showing
reasonable match between simulations and observations.
Despite the large debate that still persists on the poorly constrained
bolometric corrections and on the uncertainties on the fraction of
obscured AGN, numerical studies have shown the importance of
considering both in order to compare simulation-derived LFs and
observational ones~\cite[see, e.g., recent works
  by][]{H14,sijacki2015,OWLS2017}

Here, however, we follow a different approach.  Instead of correcting
for absorption in the SXR and HXR LFs separately, following observed
predictions, we rather assume an intrinsic absorption component for
{\it every AGN source} in the simulation, with the aim of accounting
for the effect of an obscuring torus present around the central BH.

In our implementation, the specific value of the obscurer
column-density ($N_H$) is assigned to each AGN source in a
probabilistic way, by assuming the estimated column-density
distribution of the obscurer obtained in the study
by~\cite{buchner2014} (see top-left panel of Fig.~10, in their paper)
from a sample of 350 X-ray AGN in the 4\,Ms Chandra Deep Field South.
There, the authors analyse the AGN X-ray spectra and reconstruct the
distribution of $N_H$ by using a detailed model for the obscurer,
whose geometry is suggested to be compatible with a torus with a
column density gradient, where the line-of-sight obscuration depends
on the viewing angle and the observed additional reflection originates
in denser regions of the torus.

Within the PHOX code, we include this in the construction procedure of
the X-ray emission model from AGN-like sources. In particular, an
intrinsic absorption component (at the redshift of the source) is
added to the main power-law spectrum, together with the Galactic
foreground absorption (which is the same for all the sources).  For
simplicity, we use a {\tt zwabs} absorption component (at the redshift
of the source) to represent the obscuring torus.

The complete model used to mimic the X-ray emission from AGN sources
in the simulation (i.e.\ from BH particles) therefore is:
\begin{equation}\label{mod:xspecAGN}
{\tt wabs}\times{\tt zwabs}\times {\tt zpowerlw} \,,
\end{equation}
where the first {\tt wabs} represents the foreground
Galactic-absorption component.

\section{Reconstructed SXR and HXR LFs}\label{sec:phox-lsxr-lhxr}
\begin{figure*}
\centering
\includegraphics[width=0.57\textwidth,trim=20 0 0 0,clip]{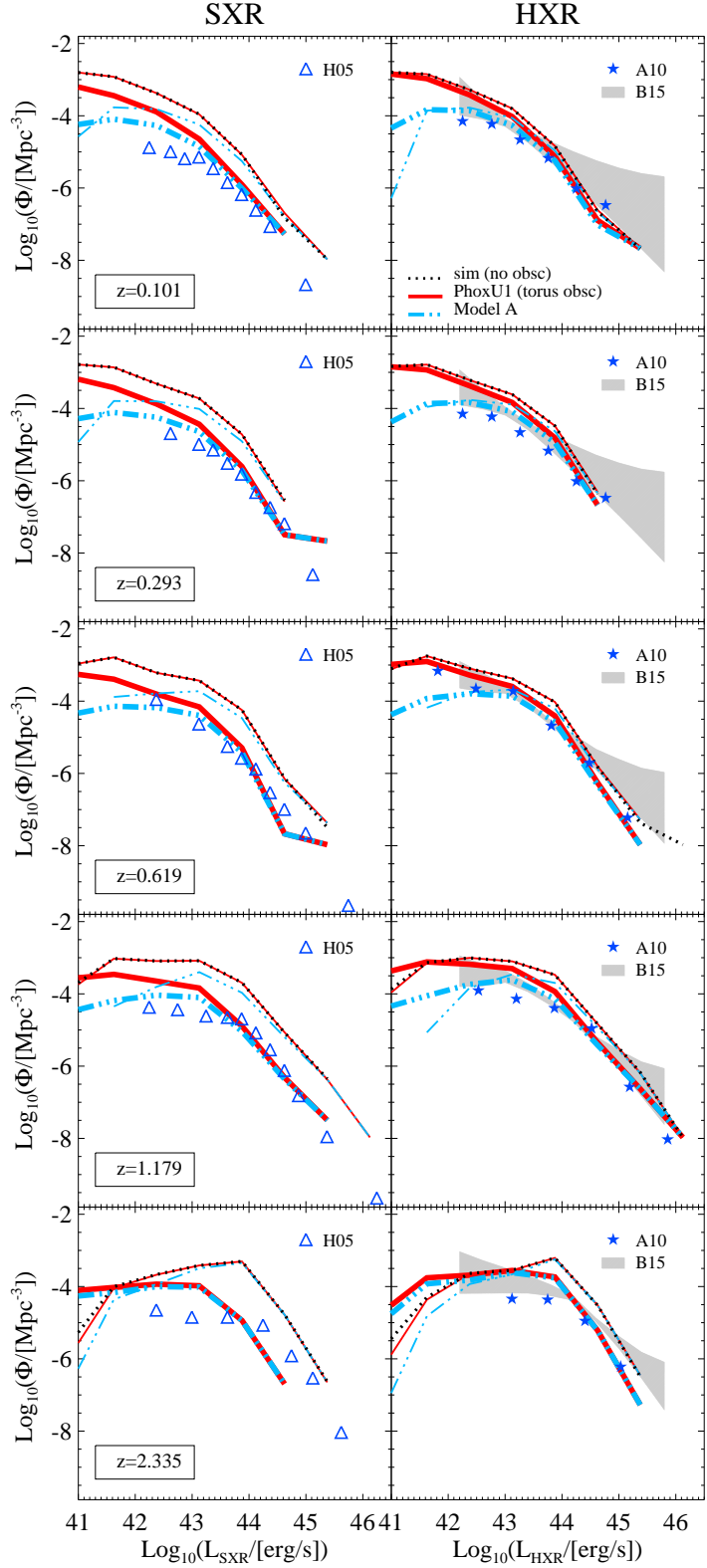}
\caption{SXR (l.h.s. column) and HXR (r.h.s. column) AGN LFs built for
  the un-absorbed luminosities from the pure simulation prediction
  (dotted black line) and for the luminosities obtained from PHOX
  Unit\,1 including the intrinsic torus absorption
  following~\protect\cite{buchner2014} (solid red lines) and the
  (soft)luminosity- and redshift-dependent obscuration (in addition to
  intrinsic torus absorption) by~\protect\cite{hasinger2008} ({\it
    Model A}; dot-dot-dashed cyan lines).  For the PHOX cases, we show
  both the LFs reconstructed from the un-absorbed intrinsic
  luminosities (thin lines) and from the absorbed ones (thick lines).
  For comparison, observed SXR and XHR LFs
  from~\protect\cite{aird2010} (blue stars),
  \protect\cite{hasinger2005} (blue triangles)
  and~\protect\cite{buchner2015} (grey shaded areas) are marked. From
  top to bottom, the panels refer to different redshifts $z\sim
  0.1,0.3,0.6,1.2,2.3$.\label{fig:sxr_hxr_lfs}}
\end{figure*}

In \fig\ref{fig:sxr_hxr_lfs} we present the results of the modelling
described in the previous subsections on the estimation of the SXR and
HXR LFs and their comparison to observational data.

In order to purely test the reliability of our treatment of AGN
emission model and intrinsic obscuration, and to discuss it in
comparison to observations, we do not construct at this stage complete
mock observations for a given X-ray instrument. Simply, we consider
the ideal photon list, generated from PHOX {\scshape unit\;1}, for each
AGN in the simulated cosmological volume.
For the purpose of this test, we set the foreground Galactic
absorption to zero, so that the ideal photons obtained from PHOX
{\scshape unit\;1} can directly be used to calculate soft and hard X-ray
LFs.
Fiducial values for collecting area and observing time are
significantly larger than realistic observational parameters, namely
$A=2e3\,\cm^2$ and $\tau=50\ks$, respectively.  This allows for
luminosity estimates that closely resemble theoretical expectations.
As assumed above, we construct the LFs for the rest-frame SXR and HXR
energy bands, i.e. we integrate the flux from the observed photon
lists within the redshifted energy bounds.\footnote{Namely,
  [$E_1$--$E_2$] and [$E_2$--$E_3$], with
  $E_1=0.5/(1+z)$\,keV,$E_2=2/(1+z)$\,keV and $E_3=10/(1+z)$\,keV, for SXR and HXR bands respectively.}

For different redshifts, from $z\sim 0.1$ up to $z\sim 2.3$, we
compare in \fig\ref{fig:sxr_hxr_lfs} the SXR and HXR LFs build from
the simulation without any absorption (dotted black curves) and
  the ones reconstructed from PHOX.

  For comparison,
  we overplot observational data presented by~\cite{aird2010},
  \cite{hasinger2005} (blue stars and triangles, respectively),
  and~\cite{buchner2015} (grey shaded areas).

  The LFs for the AGN processed with PHOX, taking into account
    the intrinsic torus absorption as in Sec.~\ref{sec:phox-obsc}, are
    shown in \fig\ref{fig:sxr_hxr_lfs} as red curves: the thick lines
    refer to the absorbed luminosities, whereas the thin lines
    indicate the LFs constructed from the intrinsic un-absorbed
    luminosities instead.

  We show both LFs from un-absorbed and absorbed luminosities to
    show the two extreme cases, considering that different datasets
    make different assumptions on this point.  In the SXR band
    especially, datasets that include only un-absorbed (type I) AGN,
    like those by~\cite{hasinger2005}, directly employ the observed
    un-corrected luminosities and should be compared against simulated
    thick curves, that are produced from PHOX absorbed luminosities
    directly. The agreement in this case is relatively good,
    especially at intermediate-high redshifts and luminosities
    $\gtrsim 10^{43}$\,erg/s.
    Other observational datasets that include instead both absorbed
    and unabsorbed AGN and construct LFs from absorption-corrected
    luminosities, as in the case of~\cite{buchner2015}, should be
    rather compared to the simulated LFs (thin lines) built from
    intrinsic luminosities.  The difference between the two curves in
    the HXR band is not striking and the simulated results show an
    overall good agreement with observations up to redshift $z\sim
    1.2$ (depsite a mild overestimation of the LFs around $L_{\rm HXR}
    \sim 10^{44}$\,erg/s).

  Interestingly, this intrinsic-absorption modelling (``PHOXU1 (torus
  obsc)'' in Fig.~\ref{fig:sxr_hxr_lfs}) naturally improves the
  comparison between simulations and observations, both in the soft
  and in the hard X-ray bands, at the same time.  In fact, while the
  simulated LFs typically over-estimate the number of AGN at all
  luminosity scales, the curves obtained from PHOX, including the
  intrinsic absorption, tend to shift towards the range occupied by
  the observed data points.
This evidence stresses the reliability of our implementation, as in
fact by assuming an observational-based distribution of intrinsic
absorption for the torus around each AGN source in the simulations
(whose scales cannot be directly probed due to resolution limits), we
can statistically better reproduce the observed LFs, not only in the
HXR band from which the observed distribution is derived, but also in
the SXR band.
In the following analysis, for the purpose of generating synthetic
observations of clusters including ICM and AGN emission, we will
therefore use for the AGN component this model including an instrinsic
torus-like absorption (``PHOXU1 (torus obsc)'' in
Fig.~\ref{fig:sxr_hxr_lfs}).

As previously mentioned,
while some observational studies comprise both absorbed
and unabsorbed AGN~\cite[][]{buchner2014,aird2015,buchner2015}, other
datasets especially in the SXR band specifically refer to
unabsorbed (type I) AGN, as for the LFs by~\cite{hasinger2005}
reported in Fig.~\ref{fig:sxr_hxr_lfs}.
 In order to compare more faithfully simulation results to this kind
 of observations, we also generate LFs from AGN processed with PHOX by
 statistically excluding some of the sources according to an
 obscuration fraction that depends on the source SXR luminosity and
 redshift, as suggested by many observational evidences~\cite[e.g.][]{ueda2003,lafranca2005,simpson2005,hasinger2005,hasinger2008,ueda2014,liu2017}.

In particular, for this additional test and with the only
purpose of comparing with data, we derive absorbed LFs from PHOX by
adopting
also the statistical obscuration fraction proposed
by~\cite{hasinger2008}, where the fraction of obscured AGN ($f_{\rm
  obsc}$), at $z < 2$, is given by:
\begin{equation}\label{fobsc}
f_{\rm obsc}(z,L_{\rm SXR}) = -0.281\left(\log(L_{\rm SXR})-43.5\right)+0.279\,(1+z)^\beta,
\end{equation}
with $\beta=0.62$ providing the best fit to their bservational data.
For higher redshifts, $z>2$, the value of $f_{\rm obsc}$ remains
approximately the same as for $z=2$:
\begin{equation}\label{fobsc.2}
f_{\rm obsc}(z,L_{\rm SXR}) = -0.281\left(\log(L_{\rm SXR})-43.5\right)+0.551.
\end{equation}

In Fig.~\ref{fig:sxr_hxr_lfs} we also report the LFs derived with this
approach, marked as {\it Model A} (cyan, dot-dot-dashed lines).
Essentially, we apply the relations~\eqref{fobsc} and~\eqref{fobsc.2}
to the intrinsic $\lsxr$, as predicted by the simulations, in order to
stochastically decide whether the AGN will be obscured or not,
depending on its luminosity and redshift.

Given that studies like those by~\cite{hasinger2005}
or~\cite{aird2010} do not explicitly correct the luminosities by the
absorption, even though they still include some moderatly obscured AGN
(with intrinsic column densities $N_H < 10^{23}$\,cm$^{-2}$), after
considering the statistical obscuration fraction only the
``observable'' AGN are processed with PHOX, and to those, despite not
being optically-thick, we still assign a torus absorption value (from
the $N_H$ distribution by~\cite{buchner2014}) to account for partial
obscuration. We note that this is an extreme limit that might
overestimate the effect of obscuration.

With respect to the LFs derived from PHOX including the torus
absorption without any explicit luminosity and redshift dependency,
the {\it Model A} produces a decrease, at all redshift, in the LFs
at the low-luminosity end ($\lsxr \lesssim 10^{43}$\,erg/s and $\lhxr
\lesssim 10^{44}$\,erg/s), where absorption effects are more severe,
whereas no significant change is observed in the high-luminosity tail.
In general, this additional modelling of the luminosity- and
redshift-dependent obscuration, combined with the intrinsic absorption
implemented within PHOX, contributes to further improve the comparison
against the observed luminosity functions by~\cite{hasinger2005}
and~\cite{aird2010}, reported in Fig.~\ref{fig:sxr_hxr_lfs}.
The improved agreement with observed LFs is particularly evident in
the low redshift bins.
Even for {\it Model A} we show both LFs for absorbed and
  intrinsic luminosities (thick and thin lines, respectively), to
  emphasize the two extremes. Like before, the most significant
  differences are visible in the SXR band, while no strong effects are
  found in the HXR. In particular, we note that the comparison against
  the observations by~\cite{hasinger2005} is more faithful when
  absorbed luminosities are employed (thick cyan curves), and in this
  case the agreement is also overall better.

In general, there are large uncertainties and differences
characterising different observational datasets, e.g. about the
definition of obscured or un-obscured AGN depending on the value of
$N_H$ or on the use of absorption-corrected or observed luminosities
to build up the LFs.
For this reason, in our implementation within PHOX, we assign a torus
obscuration component to each observed AGN (according to
the~\cite{buchner2014} distribution) but do not include any specific
luminosity or redshift dependency of the obscuration fraction.


\section{ICM and AGN X-ray emission in simulated clusters}\label{sec:results}

In this section we present the simulation results on the X-ray
emission from AGN in galaxy clusters, and its contribution to the
global X-ray emission from the ICM.

\subsection{Simulated data set}
We extract a sample of cluster-size haloes, at various cosmic times,
from the \sims\ cosmological volume of the {\it Magneticum Pathfinder
  Simulation} set.
Specifically, we select all the haloes with $\mfive > 3\times 10^{13}
\msunh$, at 9 redshift snapshots between $z\sim 0.07$ and
$z\sim2$, obtaining therefore catalogs which are complete both in mass
and volume.
At $z=0.07$ our sample comprises 1649 clusters with
masses $3\times 10^{13} \lesssim \mfive/[\msunh] \lesssim 1\times 10^{15}$,
whereas at $z=1.98$ the sample includes 34 systems with $\mfive$
spanning the mass range $3\mbox{--}6.5 \times 10^{13}$.

\subsection{Mock X-ray observations}\label{sec:eRosita-mocks}
In order to predict the relative importance of AGN emission with
respect to the ICM X-ray luminosity, we generate synthetic eROSITA
observations out of the clusters in the simulation.
Specifically, we apply PHOX to the catalogs of simulated clusters,
considering both the ICM and the AGN components in the clusters.
Ideal photon lists have been then generated for (i) the ICM X-ray
emission (as outlined in Sec.~\ref{sec:phox-icm}) and (ii) the
combined emission from ICM and AGN within the clusters (as outlined in
Sections~\ref{sec:phox-agn} and~\ref{sec:phox-obsc}).
For the AGN emission, we only include the modelling of the intrinsic
torus absorption according to the absorber column-density distribution
by~\cite{buchner2014}, as in \eq\eqref{mod:xspecAGN} (i.e.\ we model
both Compton-thick and unabsorbed AGNs --- but do not include any
explicit luminosity- and redshift-dependent obscuration fraction).
By means of the SIXTE\footnote{{\tt
    http://www.sternwarte.uni-erlangen.de/research/sixte/}.} dedicated
simulator, we subsequently convolved the PHOX ideal photon lists with
eROSITA instrumental specifications, in order to obtain
observational-like data files.  The ICM-only and
  AGN+ICM eROSITA-like images obtained with SIXTE are shown in
  Appendix~\ref{app:mocks} for two example clusters in our simulations.
Since we do not include instrumental background, the received
  count rate effectively does not depend on the exposure time assumed
(here we use 10\,ks).
eROSITA images and relative spectra have been extracted from the
region enclosed within the projected $\rfive$ for every cluster in the
catalogs, at various redshift. This set up would correspond to
simulating eROSITA pointed observations.

In this procedure, we assume for both source types a \wabs\ model in
\xspec\ to include an artificial foreground Galactic absorption, with
the value of the column density fixed to $N_H = 10^{20}$\,cm$^{-2}$.

\subsection{AGN-to-ICM X-ray luminosity}\label{sec:erosita}

Before inspecting the synthetic X-ray data, we investigate directly
the simulations to predict the relation between the X-ray
luminosity of the ICM and of the central AGN.

\begin{figure}
  \centering
  \includegraphics[width=0.49\textwidth,trim=5 0 0 0,clip]{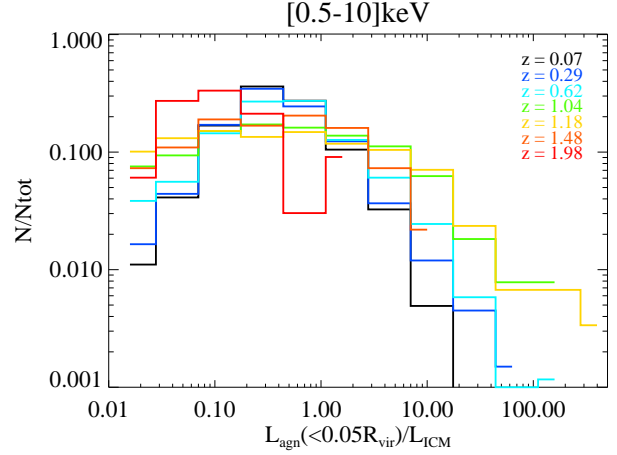}
\caption{Differential distribution function of the AGN-to-ICM
  luminosities for all the central AGN sources of the selected
    clusters, at different redshifts between $z=0.07$ and $z\sim
    2$. The AGN luminosity is the sum of the intrinsic
    luminosities (computed as in Section~\protect\ref{sec:agnsxrhxr},
    without considering any absorption) of all AGN sources residing
    within $0.05\,R_{\rm vir}$, whereas the ICM luminosity is computed
    from the gas enclosed within the virial radius.  Both luminosities
    are theoretical estimates calculated directly from the simulations
    in the [$0.5$--$10$]\,keV band.
  \label{fig:Lagn_Licm_sim}}
\end{figure}

In Fig.~\ref{fig:Lagn_Licm_sim}, we show simulation results for the
AGN-to-ICM luminosity ratio in the $[0.5$--$10]$\,keV band, at various
redshifts between $z\sim 0.1$ and $z\sim2$.  In particular, we
consider the global ICM luminosity coming from the region within the
virial radius and compare it to the intrinsic luminosity expected from
the central AGN.  To this scope we consider all AGN sources residing
within the centremost region, i.e.\ $<5\%$ of the cluster virial
radius, so that the $L_{\rm AGN}(<0.05R_{\rm vir})$ is in fact the
total sum of their luminosities. We note, nonetheless, that for the
majority of our clusters only the central AGN is comprised in that
very central region.
In order to estimate the intrinsic contribution predicted by the
simulations, here
  luminosities are the theoretical values computed directly from the
  simulations (as in Section~\ref{sec:agnsxrhxr}), so no
  absorption is included in the AGN emission.
The effects of absorption and of the observational approach will be
investigated through the mock analysis in Section~\ref{sec:res-mocks}.

We see from Fig.~\ref{fig:Lagn_Licm_sim} that the distribution of the
ratio
\begin{equation}\label{eq:L-ratio-sim}
    f_L \equiv L_{\rm AGN}(<0.05R_{\rm vir})/L_{\rm ICM}(<R_{\rm vir})
\end{equation}
is typically centered on values comprised between $0.1$ and $1$.
Up to $z\sim 0.6$ the median of the distribution slightly increases
and then oscillates around similar values up to $z\sim 1$. The mean
value, instead, increases from $z=0.07$ up to $z\sim 1.2$ and presents values even larger than 1 at redshift $0.6 \lesssim z \lesssim 1.2$.
At redshifts $z\gtrsim 0.6$ the distribution becomes broader and
presents a prominent tail at values of $f_L$ much larger than 2.
The broader shape of the distribution is still visible at $z\sim 1.5$,
although the tail is truncated at $f_L \sim 10$.
Instead, at $z\sim 2$, $f_L$ is again typically always lower than $\sim 1$.
At that redshift, however, the cluster sample is much smaller and the
AGN LFs indicate a poorer agreement between simulated and
observational results.
We cannot therefore derive strong constraints from these results at
$z\sim 2$.

These findings indicate that considering the centremost AGN source(s)
only, their intrinsic luminosity in the SXR+HXR band is typically
smaller than the luminosity of the whole ICM within the virial region
of their hosting cluster. Nevertheless, it also indicates that some
clusters do host powerful AGN emission in their core, which can easily
reach or even exceeds by several times the global ICM X-ray
luminosity.  This effect is more severe at redshifts $z\sim 1$--$1.2$,
where up to $\sim 18\%$($\sim 13\%$) of the clusters considered show
an intrinsic value of $f_L \gtrsim 2$.

\subsubsection{Mock eROSITA observations}\label{sec:res-mocks}

Given the prediction from the intrinsic luminosities, it is very
important to investigate the {\it observed} AGN contamination when
absorption and instrumental procedures are taken into account. This is
especially crucial for high-$z$ clusters, for which is more difficult
to spatially resolve clusters and AGN sources, and to disentangle the
AGN emission from that of the diffuse ICM.  We inspect this case by
means of the eROSITA synthetic spectra generated. This allows us to
directly compare the counts that one would obtain from the sole ICM
emission against the combined counts from ICM and AGN, from within the
cluster $\rfive$ radius, at various redshifts up to $z\sim
1.5\mbox{--}2$.
The AGN considered here are therefore all the sources enclosed within
$\rfive$, mimicking the worst case where individual sources cannot be
disentangled spatially.

In fact, the $\rfive$ extent of the clusters in the simulated catalogs
ranges between 80--700\,arcsec at $z\sim 0.1$ and 30--80\,arcsec at
$z\sim 1.5\mbox{--}2$. Considering that the X-ray ICM emission steeply
decreases towards the outskirts and most of it is associated therefore
to the core region, the extent of the X-ray diffuse emission for our
clusters at high redshifts is essentially comparable to the eROSITA
HEW (i.e.\ 28 arcsec in scanning mode and 15 arcsec on-axis), and
distinguishing between point sources and core ICM emission becomes
very challenging.

We investigate the expected eROSITA photon fluxes from ICM
and AGN emission in our sample, in order to constrain the relative
contribution for a 10\,ks exposure.
We compare the counts from the ICM emission with those by both AGN and
ICM, extracted by construction from the composite event file.
\begin{figure*}
\centering
\includegraphics[width=0.49\textwidth]{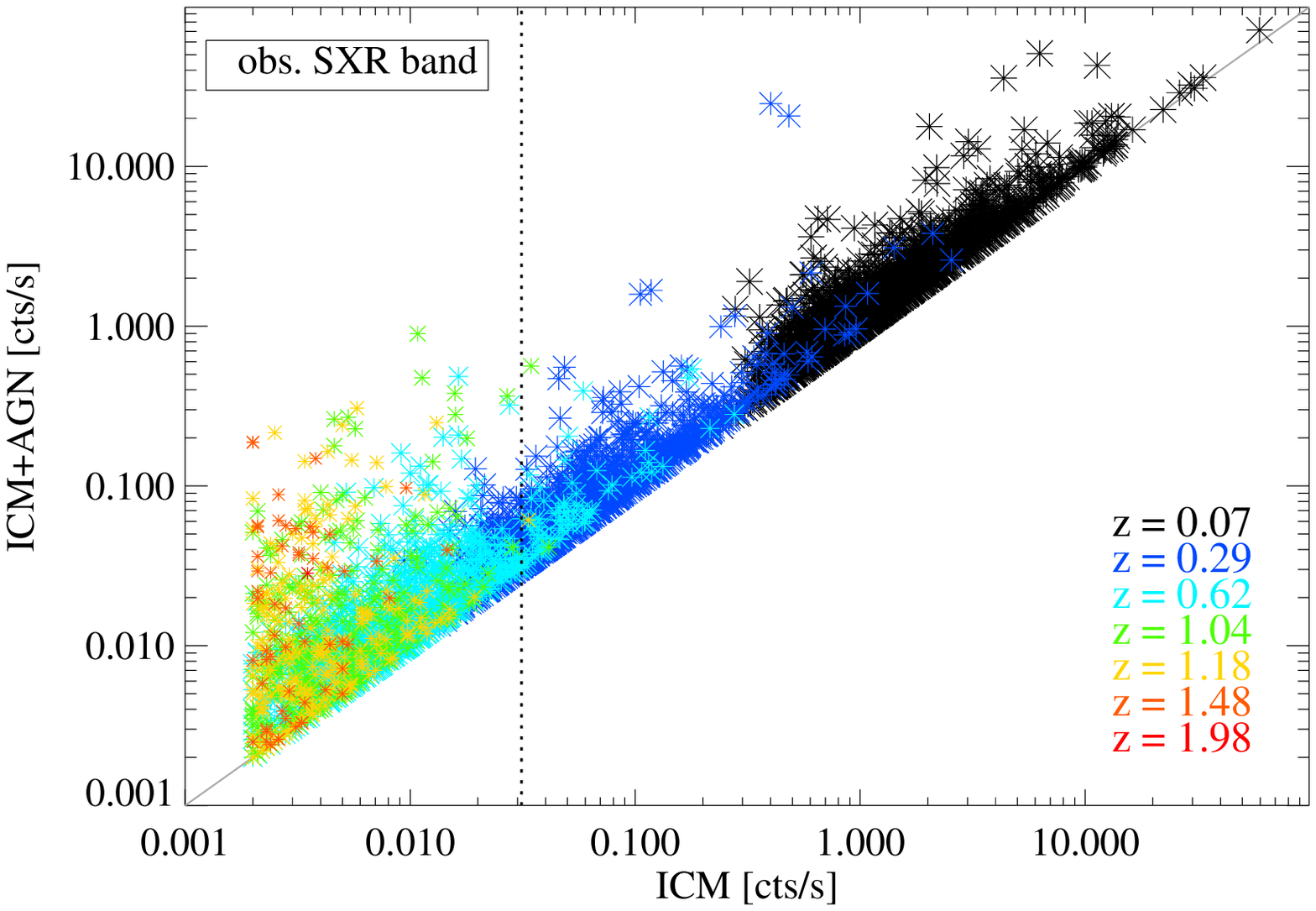}
\includegraphics[width=0.49\textwidth]{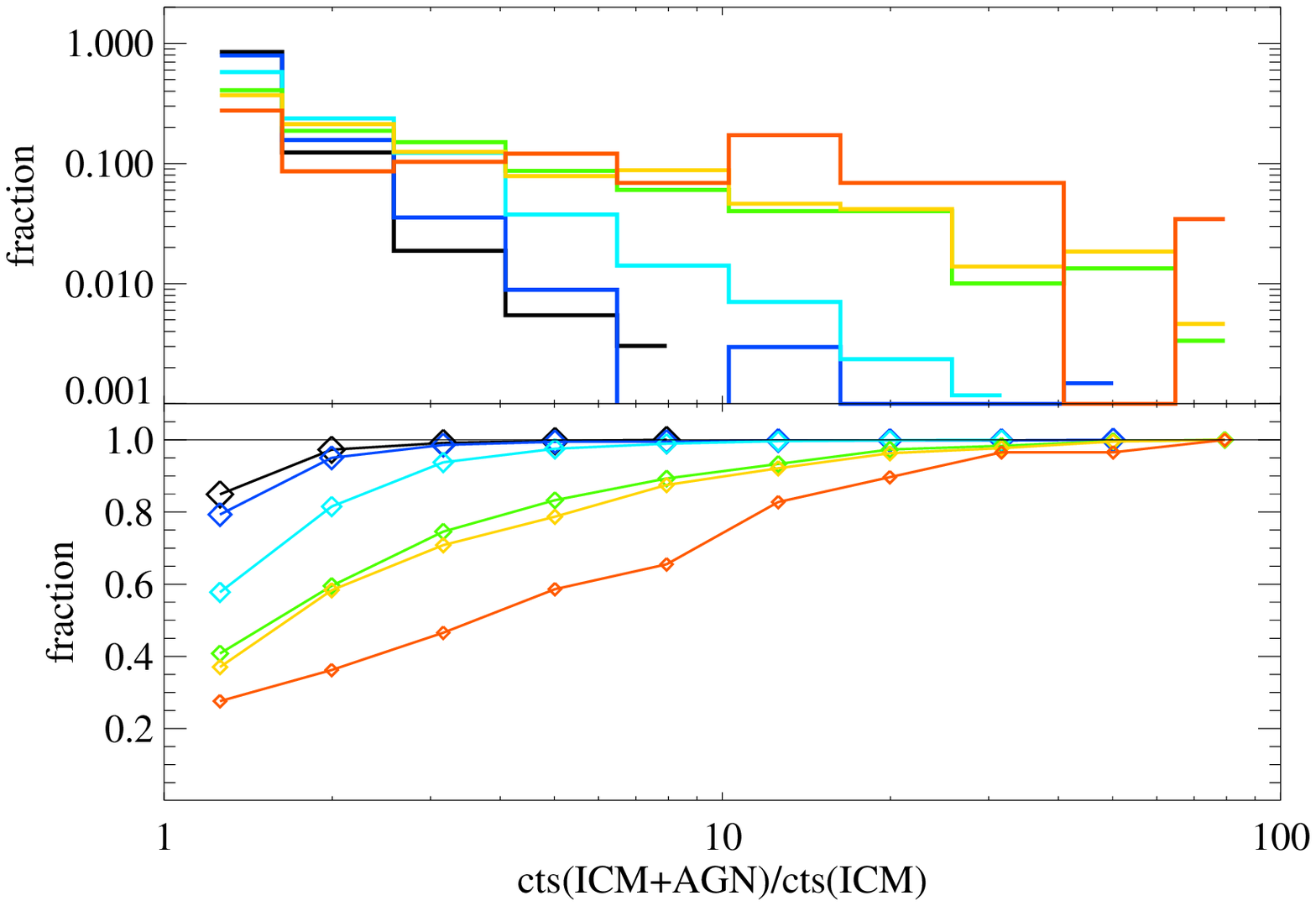}
\includegraphics[width=0.49\textwidth]{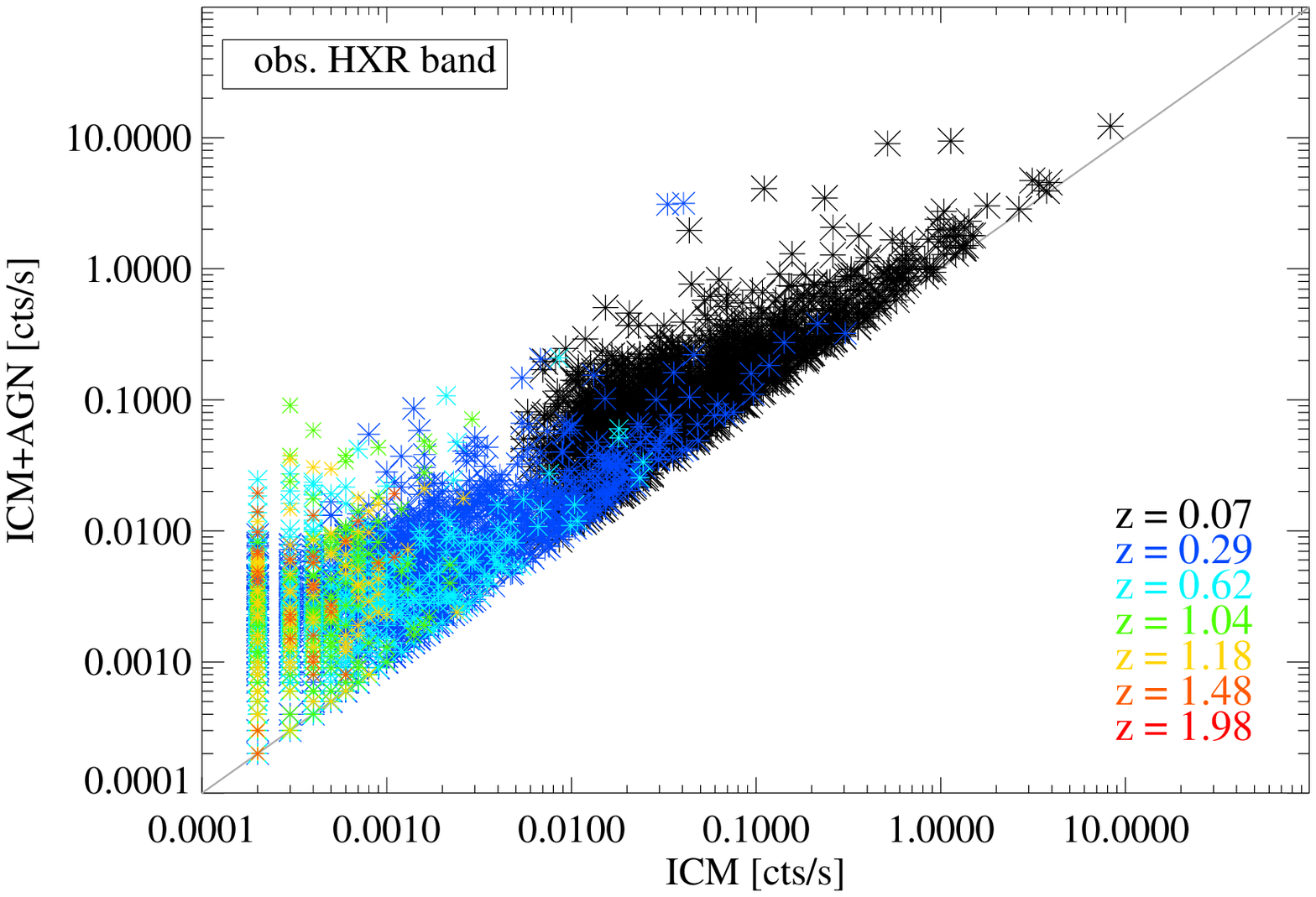}
\includegraphics[width=0.49\textwidth]{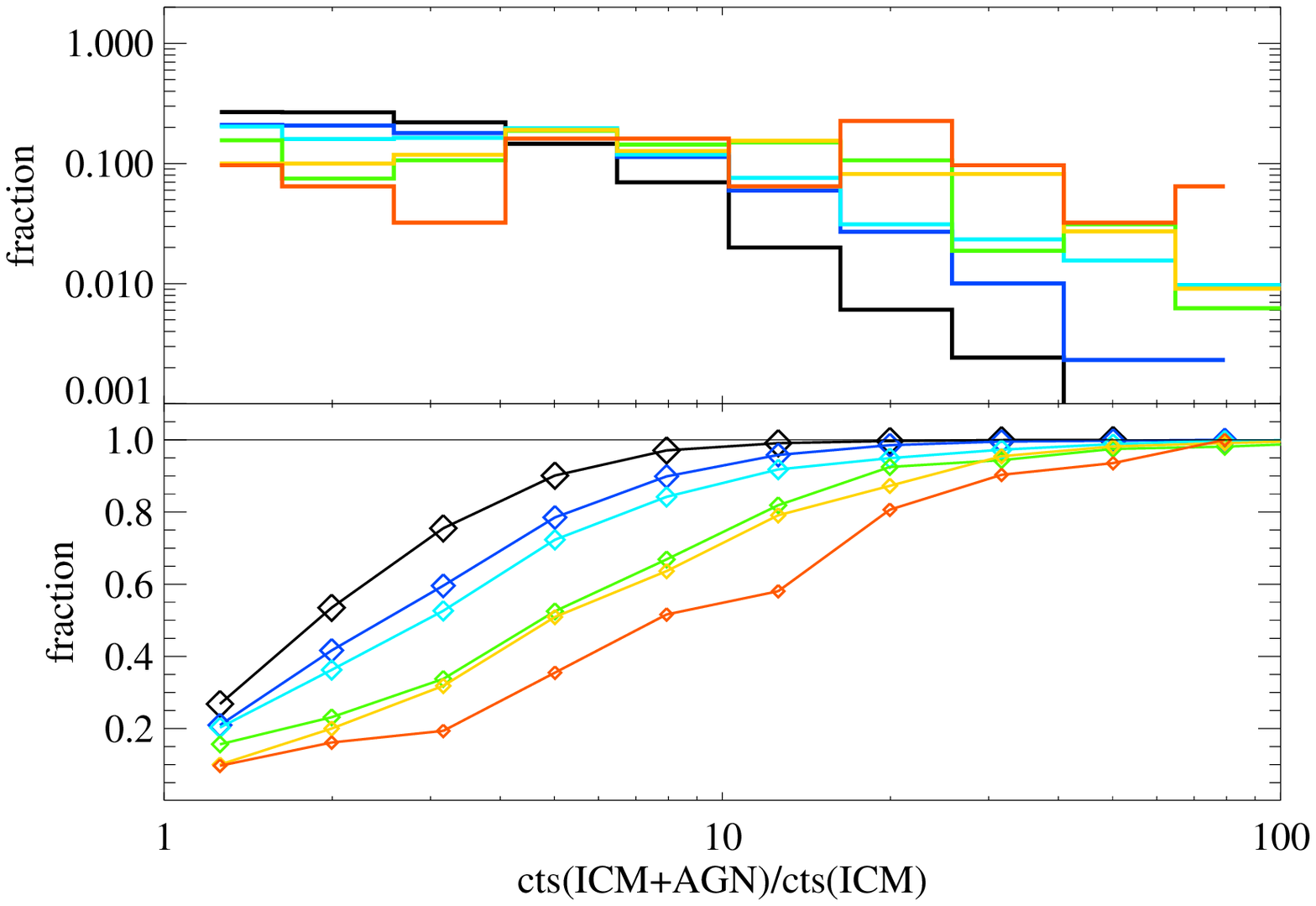}
\includegraphics[width=0.49\textwidth]{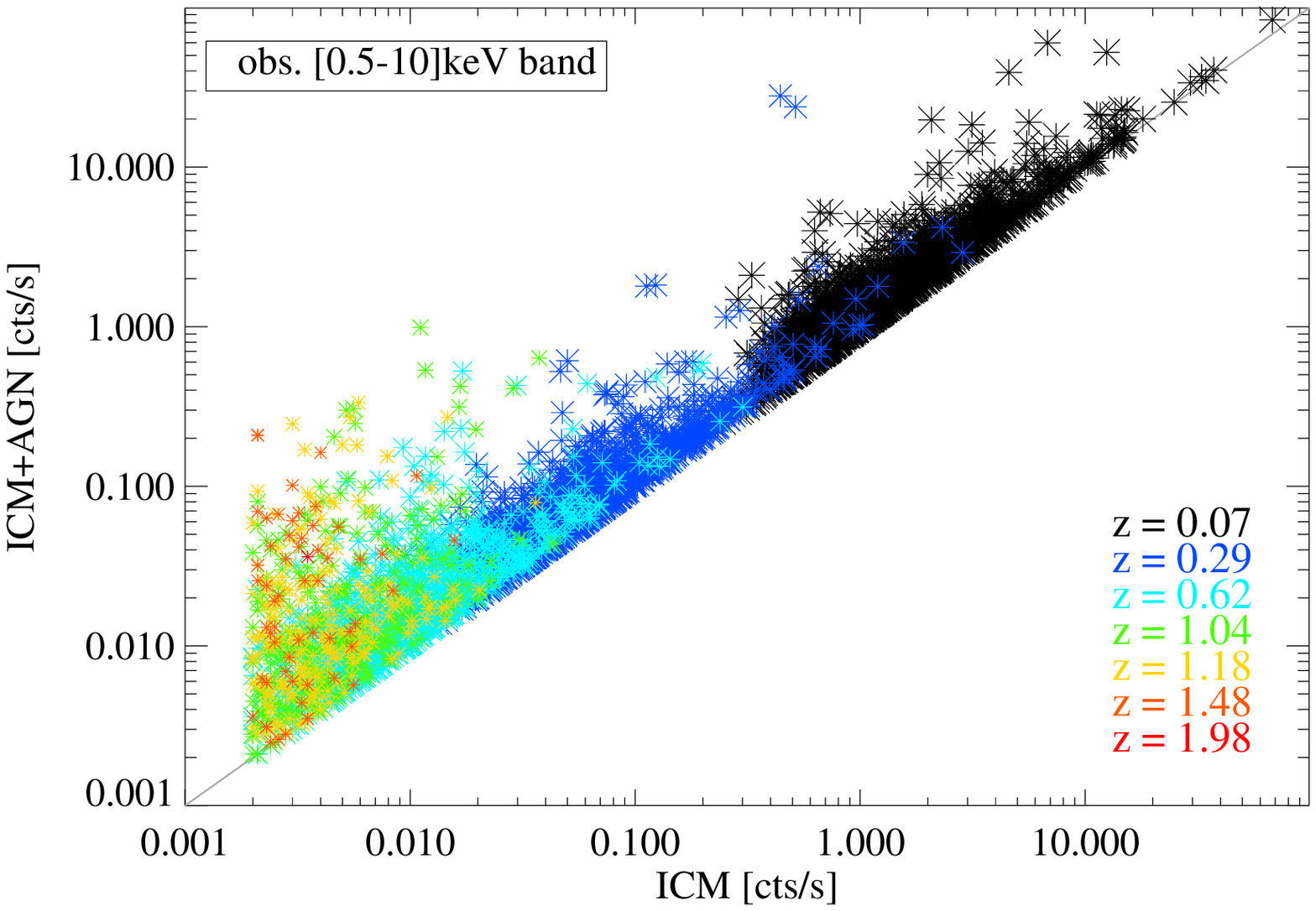}
\includegraphics[width=0.49\textwidth]{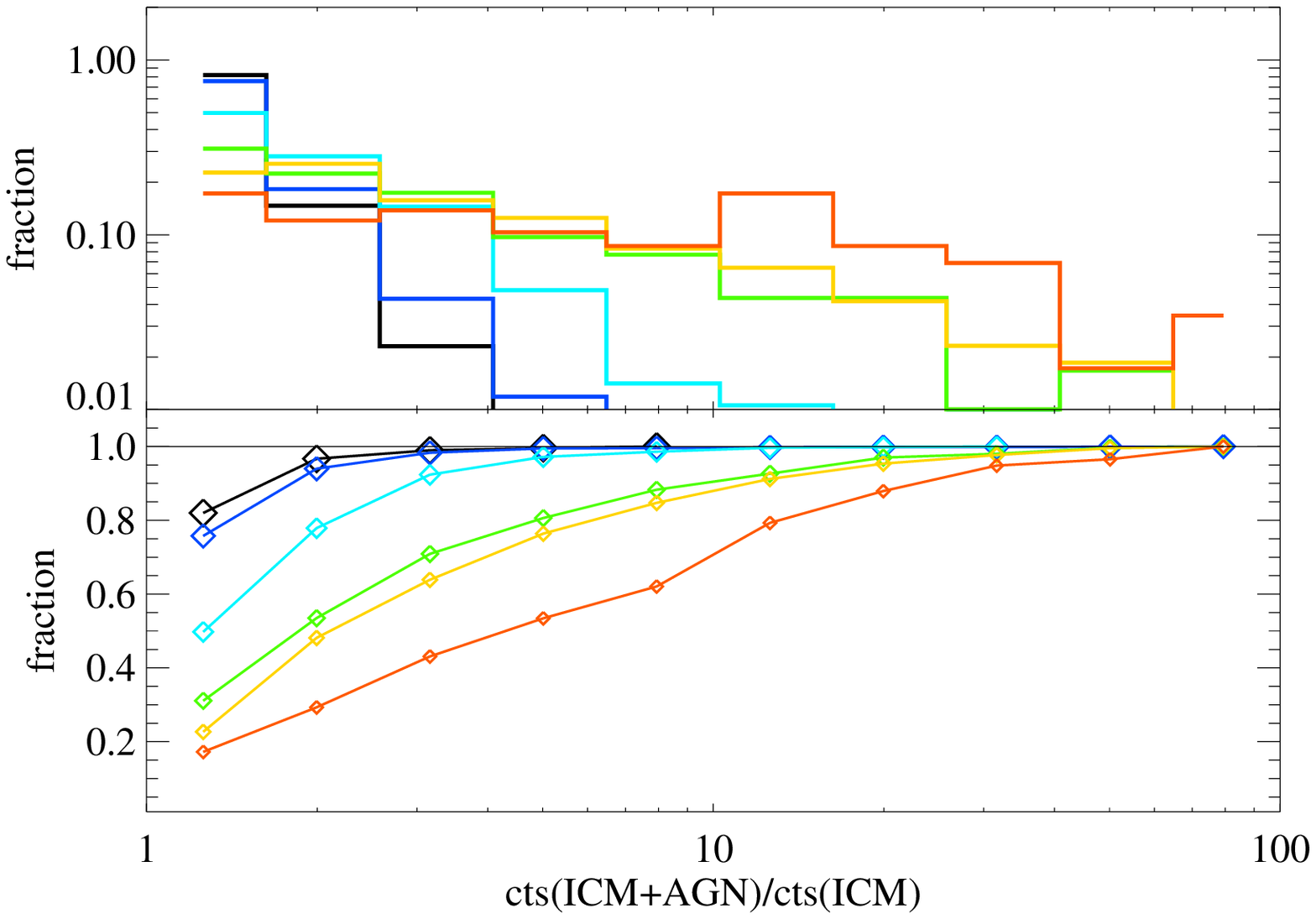}
\caption{Count rates for ICM and ICM+AGN from eROSITA mock
  observations of the R500 region of clusters at various redshifts
  between $z=0.07$ and $\sim 2$, marked with different colours as
  specified in the legend.
  {\it Left:} AGN+ICM count rate as a function of the ICM count
  rate. Each data point represents one cluster. The vertical line in
  the SXR plot marks the minimum detection threshold of 50 photons for
  a 1.6\,ks exposure.
  {\it Right:} differential (upper panel) and cumulative (lower
    panel) distribution of the ${\rm cts}(\mbox{ICM+AGN})/{\rm
      cts}(\mbox{ICM})$ ratio (see Eq.~\protect\eqref{eq:L-ratio-cts})
    for the clusters at each redshift.  Distributions are reported
    only when the sample includes more than 3 sources.  Colours are
    the same as in the l.h.s.\ panels and the distributions are
    normalized to the total number of sources at each redshift.
  From top to bottom, the three panels refer to the {\it observed}
  SXR, HXR and [0.5--10]\,keV energy bands, and only sources with a
  minimum of 20 (ICM) photons in the SXR (for the assumed 10\,ks exposure)
  are considered.
  \label{fig:cts_agn_gas_erosita}}
\end{figure*}
\begin{figure*}
  \centering
  \includegraphics[width=0.49\textwidth,trim=0 0 0 15,clip]{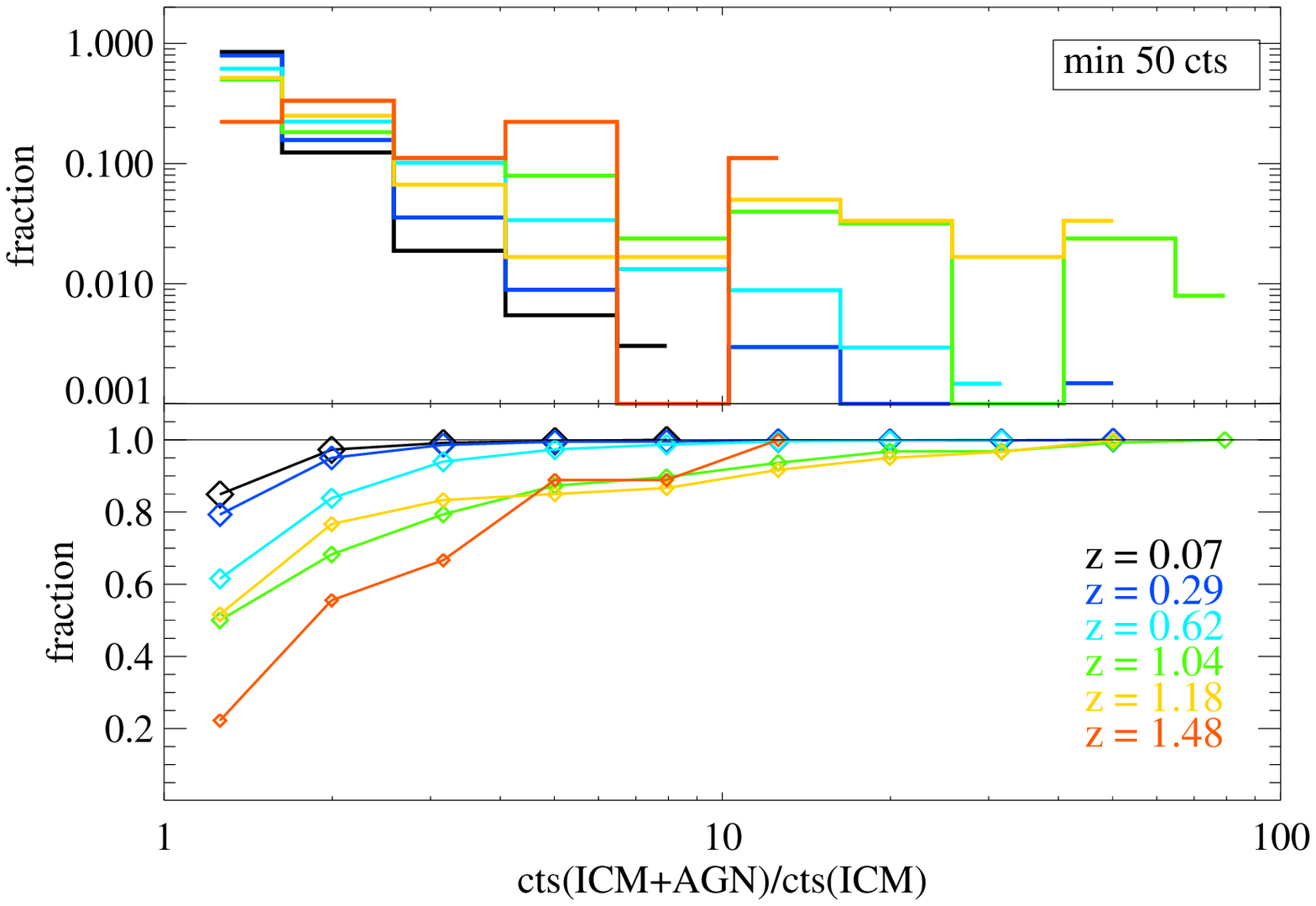}
  \includegraphics[width=0.49\textwidth,trim=0 0 0 15,clip]{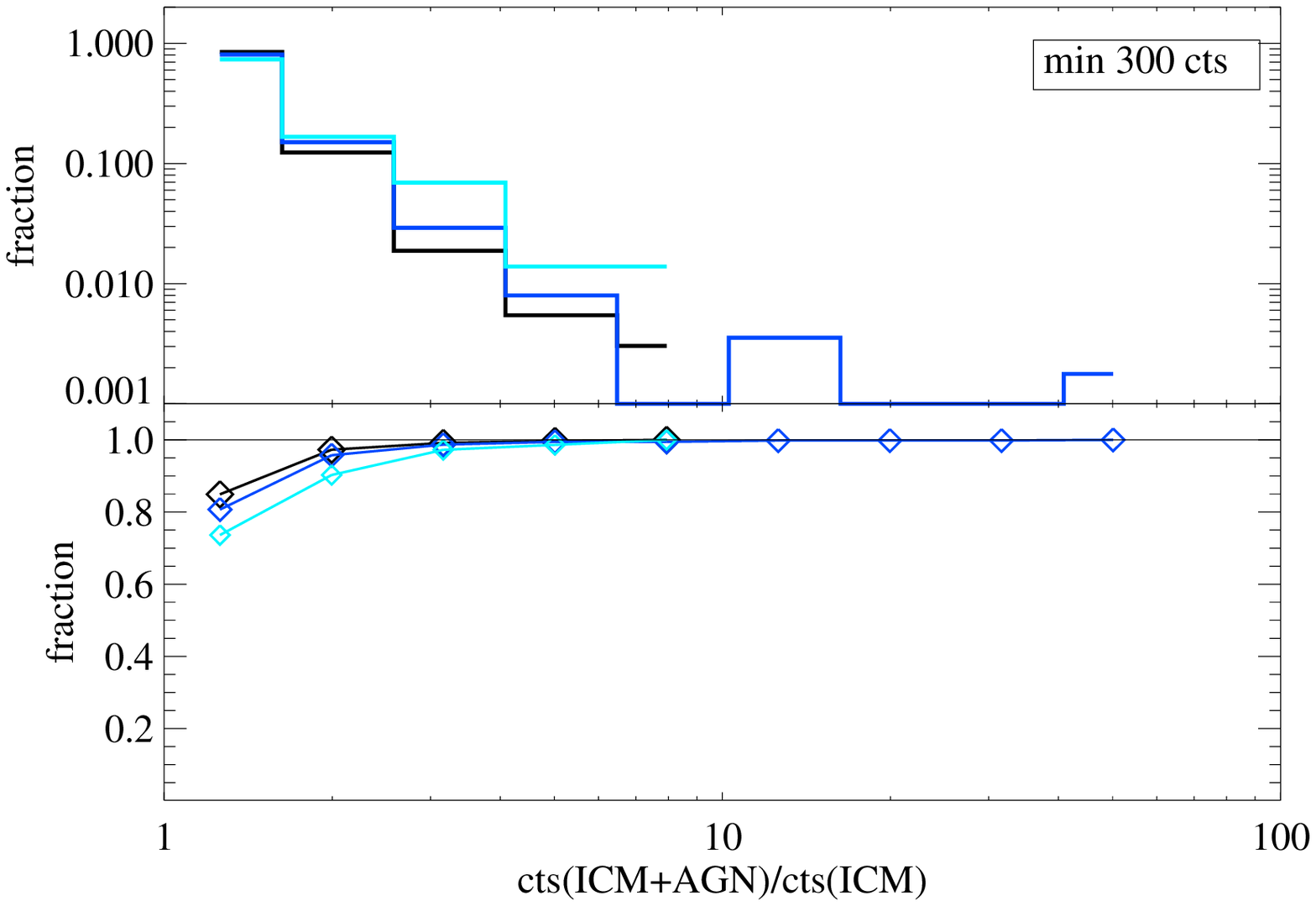}
  \caption{Same as the top-right panel in
    Fig.~\ref{fig:cts_agn_gas_erosita} (SXR), for
    different thresholds on the minimum ICM counts received in the SXR
    band: $50$ (left) and $300$ (right) photons, for the assumed
    10\,ks exposure.\label{fig:cts_sxr_thrsld}}
\end{figure*}
These results for the {\it observed} SXR, HXR and [0.5--10]\,keV energy
bands are shown in Fig.~\ref{fig:cts_agn_gas_erosita}, for the
simulated clusters catalogs at redshifts comprised between $z\sim 0$
and $z\sim 2$.

In the l.h.s.\ panels we show the distribution of the observed counts
per second due to the sum of ICM and AGN emission versus the count rate
due to the ICM only. Different colors refer to different redshifts,
as in the legend.
In the r.h.s.\ panels, we report a more quantitative
representation of the ratio
\begin{equation}\label{eq:L-ratio-cts}
  f_{\rm cts} \equiv {\rm cts}(\mbox{ICM+AGN})/{\rm cts}(\mbox{ICM}),
\end{equation}
  in terms of differential (histograms; upper insets) and cumulative
  distribution (asterisk symbols; lower insets).

From the inspection of the upper-row panels in
Fig.~\ref{fig:cts_agn_gas_erosita}, relative to the observed SXR band,
we note a trend with redshift such that the number of sources with
${\rm cts}(\mbox{ICM+AGN})/{\rm cts}(\mbox{ICM}) \gtrsim 2$ (i.e. where the
photon flux due to AGN is comparable to or larger than the ICM one)
increases from a few percents at $z=0.07$, to 30\% at $z\sim0.6$, and to
a maximum of $\sim 70\%$ at $z\sim 1.5$.  This indicates that in the
typical redshift-range of eROSITA observations ($z\sim
0.2\mbox{--}0.4$) only about 10--15\% of the clusters will be dominated
by AGN emission.

In the range $z\sim 1$--$1.2$ roughly 20\%(10\%) of the simulated clusters
show a number of counts from AGN and ICM that is at least 5(10) times
higher than that from ICM only.
At high redshift, $z\sim 1.5$, this effect is very severe, and we find
that, statistically, the X-ray flux coming from AGN is 10 times higher
than the ICM flux for $\sim 34$\% of the clusters.

The vertical line in the SXR plot marks the minimum detection
threshold of 50 photons for a 1.6\,ks
exposure~\cite[][]{pillepich2012}, which corresponds approximately to
the expected detection threshold for clusters in the eROSITA all-sky
survey. With this more stringent limit, we note that only 3 sources
from our simulation box would be detected at $z\sim1$--$1.2$, and zero
sources at $z\sim 1.5$, if the limit is applied to
the ICM flux only. Nevertheless, the photons received from the source would
comprise as well those emitted by the AGN in the cluster. If we
therefore apply the same cut to the $y$-axis instead, i.e.\ to the
AGN+ICM count rate, then one would detect 77 sources at $z\sim
1$--$1.2$ and 30 sources at $z\sim 1.5$. This means that almost all of
them are actually dominated by the AGN emission.

In the central- and bottom-row panels of
Fig.~\ref{fig:cts_agn_gas_erosita} we report analogous results for the
observed HXR and [0.5--10]\,keV bands, where the distributions are shifting
towards higher values of the $f_{\rm cts}$ ratio for increasing
redshifts.
The main difference is shown by the HXR case, where the power-law-like
AGN emission tends to dominate over the thermal ICM one.  By
considering only clusters with at least 20 ICM counts in the SXR band,
we note that for the faintest sources the corresponding count rates in
the observed HXR can be one order of magnitude lower, that is few
photons (or no photons at all) are received even in the long 10\,ks
exposure considered here.
The fraction of sources (with at least 20 ICM counts in the SXR band)
that are not detected at all in the HXR band increases with redshift,
from less than $1\%$ at $z\sim 0.07$ up to $\sim56\%$ at $z\sim
1.5$. (The only source at $z\sim 2$ is not detected in the HXR band.)
With our requirements for the detection in the SXR band, much less
stringent than that indicated by~\cite{pillepich2012}, the HXR count
rates of the observed sources indicate a contamination by AGN that is
always more severe than in the SXR or [0.5--10]\,keV bands.
In the HXR band as well, clusters can be still
detected up to $z\sim 1.5$, where however $\sim 90\%(70\%)$ of the
sources have a dominant AGN flux with respect to the ICM emission,
$f_{\rm cts}\gtrsim 2(5)$.  Interestingly, and differently from the
other bands, even at $z\sim 0.07$, the photon flux due to AGN is
comparable to or larger than the ICM one (i.e.\ $f_{\rm cts}\gtrsim
2$) already for $\sim 40\%$ of the clusters.

This is not the case for the distributions in the observed SXR and
[0.5--10]\,keV bands, where instead the distribution is narrower at
low redshifts ($z< 0.6$) and strongly peaked around $f_{\rm cts} \sim
1$--$2$, whereas it significantly broadens towards higher values of
$f_{\rm cts}$ with increasing redshift, and almost flattens at
$z\sim1.5$.

At redshifts $z\sim1$--$1.5$ we therefore expect eROSITA to encounter
a significant number of sources (10--30\%) for which the detection of
the cluster around the AGN is very challenging, with the observed
X-ray flux from the latter dominating over the former by more than a
factor of 10.

\subsubsection{AGN-to-ICM contamination and system mass}\label{sec:res-mocks-mass}

In Fig.~\ref{fig:cts_sxr_thrsld} we show the distribution of $f_{\rm
  cts}$ for two different lower cuts on the minimum of ICM photons
received in the observed SXR energy band, namely $50$ (l.h.s.\ panel)
and $300$ (r.h.s.\ panel) photons for the $10$\,ks exposure.

\begin{figure}
  \centering\includegraphics[width=0.49\textwidth,trim=0 0 0 15,clip]{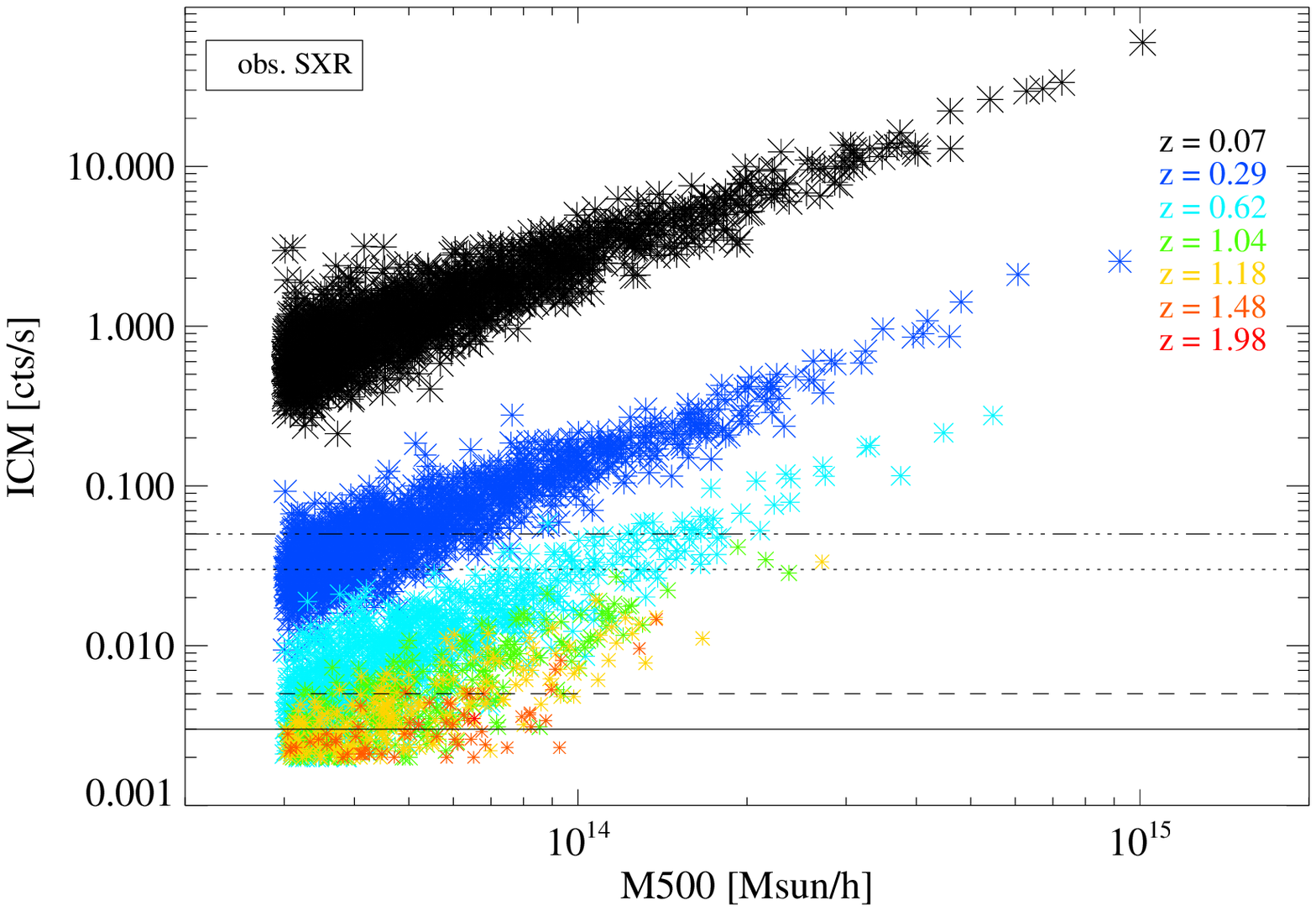}
\caption{ICM count rate in the SXR band as a function of the cluster
  mass $\mfive$, at various redshifts between $z=0.07$ and $\sim 2$,
  marked with different colours as specified in the legend. The
  horizontal lines correspond, from bottom to top, to detection limits
  of $30$ (solid line), $50$ (dashed line), $300$ (dotted line) and
  $500$ (dot-dashed line) photons for a $10\,$ks
  exposure.\label{fig:cts_sxr_m500}}
\end{figure}

As shown in the r.h.s. panel of Fig.~\ref{fig:cts_sxr_thrsld}, we
expect to detect clusters with $\mfive > 3\times 10^{13}\msunh$ only
up to redshift $z\sim 0.62$, when a detection limit of 300 photons is
applied (corresponding to the limit of 50 photons for a typical
1.6\,ks exposure of the planned eROSITA all-sky survey). At
low-intermediate redshifts ($z \lesssim 0.6$) we still find that for
the majority of the sources the AGN emission is at most comparable to
that of the ICM alone, and no more than 20\% of the sources have a
dominant AGN emission.

For less stringent limits on the detection limit, such as 50 counts in
10\,ks (l.h.s.\ panel in Fig.~\ref{fig:cts_sxr_thrsld}), also fainter
sources can be observed in the SXR band and up to $z\sim 1.5$. In this
case, a larger fraction of sources is dominated by the AGN
emission, namely up to 30--60\% of the clusters would have $f_{\rm
  cts}>2$ between redshifts $1$ and $1.5$.

By increasing the threshold on the SXR count rate detection we
essentially aim at increasing the mass limit of the clusters that can
be detected and obtain that only the brightest sources can be detected
at high redshifts.  Nevertheless, we note from
Fig.~\ref{fig:cts_sxr_m500} that the received ICM count rate in the
SXR band correlates tightly with the system mass (here, $\mfive$) only
for the massive clusters. At lower masses, at fixed values of ${\rm
  cts(ICM)}$ corresponds in fact a large scatter in $\mfive$, for all
the redshift considered.
In addition, given the trend in redshift of the ${\rm cts(ICM)}$-$\mfive$
relation shown in Fig.~\ref{fig:cts_sxr_m500}, we note that the
increasing threshold on the minimum ${\rm cts(ICM)}$ (horizontal lines in
the Figure, corresponding to detection limits of $30$, $50$, $300$ and
$500$ photons in $10\,$ks exposure, from bottom to top respectively)
corresponds to an increasing mass cut of the observed clusters only at
high redshifts.  Instead, for $z\lesssim 0.3$, the entire mass range
of the sample $\mfive \gtrsim 3\times 10^{13}M_\odot$ is still
observable, even though some sources with low mass and low count rate
cannot be detected \cite[for a discussion of the observational effects of this scatter, see e.g.][]{allen2011}.
We remind that Fig.~\ref{fig:cts_sxr_m500} includes all the sources
with at least 20 ICM counts detected in the SXR band, as in
Fig.~\ref{fig:cts_agn_gas_erosita}. We verified nonetheless that the
presence of sources that are dominated by the AGN emission --- for
which the correct estimation of the ICM count rates can be therefore
compromised --- does not introduce any bias. By removing the
AGN-dominated sources, in fact, the number of sources is reduced with
a more prominent effect at high redshift, but the trends observed in
Fig.~\ref{fig:cts_sxr_m500} are preserved, making our conclusions
unchanged.

\begin{figure}
  \centering\includegraphics[width=0.49\textwidth,trim=0 0 0 15,clip]{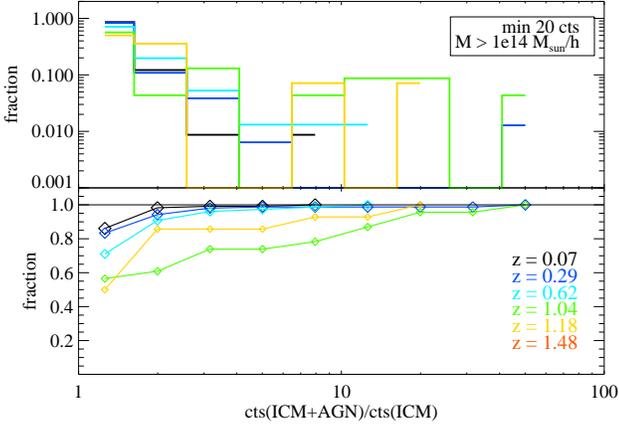}
\caption{Same as the top-right panel in
  Fig.~\ref{fig:cts_agn_gas_erosita} (SXR energy band), for the
  cluster in the sample with $\mfive \gtrsim 10^{14}M_\odot$. As in
  Fig.~\ref{fig:cts_agn_gas_erosita}, here only sources with at least
  20 counts in the SXR band (for the 10\,ks exposure) are
  considered.\label{fig:cts_sxr_mcut}}
\end{figure}

For the purpose of comparison, we show in Fig.~\ref{fig:cts_sxr_mcut}
the distribution of $f_{\rm cts}$, in the observed SXR band, for the
subsample of clusters with $\mfive \gtrsim 10^{14}M_\odot$. (As in
Fig.~\ref{fig:cts_agn_gas_erosita}, we require a minimum of 20 ICM
photons in the SXR band, for the assumed 10\,ks exposure.)  With
respect to Fig.~\ref{fig:cts_sxr_thrsld}, we see that the most massive
clusters can be detected up to redshift $z\sim 1.2$, with a small
contamination due to AGN in the majority of the cases. Nevertheless,
at $z\sim 1\mbox{--}1.2$ the $f_{\rm cts}$ distribution shows a tail
towards higher values, due to the large scatter, at fixed mass, in the
AGN emission with respect to the ICM one. From this case, we infer
that even for massive clusters, $15$--$25$\% of the sources at $z\sim
1\mbox{--}1.2$ can be largely dominated by their AGN component in the
observed SXR photon flux, with an AGN count rate that is at least 5
times larger than the ICM count rate ($f_{\rm cts} \gtrsim 5$).

\section{Summary and conclusion}\label{sec:conclusion}

In this paper we have presented a new implementation of the X-ray
emission modelling from AGN sources within the {\it Magneticum
  Pathfinder Simulation} set.  We showed that the population of AGN
in the {\it Magneticum Pathfinder Simulation} statistically well
reproduces the observed bolometric unabsorbed and SXR/HXR absorbed LFs
up to redshift $z\sim 2$, over a large range of luminosities.  As an
application, we presented predictions on the eROSITA observations of
AGN and ICM X-ray fluxes for a sample of galaxy clusters, in the range
$0 \lesssim z \lesssim 2$, quantifying the contamination due to the
AGN emission w.r.t.\ the ICM X-ray flux.

Our main findings can be summarized as follows.

Bolometric LFs from the simulated AGN catalogs have been analysed and
compared against observational data in order to evaluate the
reliability of the intrinsic simulated population of SMBHs
(Fig.~\ref{fig:bol_xlf})~\cite[see also][]{H14}. The theoretical
estimation of their bolometric luminosity and the therefrom derived
unabsorbed SXR and HXR luminosity (under the assumtion of the
bolometric corrections by~\cite{marconi2004}) were then used to
constrain {\it in a self-consistent way} the spectral parameters of
the X-ray emission associated to each AGN source. In the modelling of
the X-ray emission from each source, we further statistically
associate to each AGN source a value for the column density of the
obscuring material expected to reside in the torus surrounding the
AGN, by assuming an observationally-motivated distribution of the
obscurer column densities~\cite[from][]{buchner2014}.  This modelling
is embedded into our PHOX photon simulator in order to derive the
X-ray synthetic emission from the AGN component in the simulations
(Section~\ref{sec:phox-agn} and~\ref{sec:phox-obsc}).
The assumption of an obscuration component at the sources allowed us
to test the obscured LFs in both the soft and hard X-ray energy bands
(Section~\ref{sec:phox-lsxr-lhxr}), which show a remarkable level of
agreement with observational LFs for the high-luminosity tail in all
the redshift bins analysed between $z=0.1$ and $z=1.2$.  The simulated
LFs still overpredict the observational ones at low luminosities, in
both energy bands.  Nevertheless, when an additional luminosity- and
redshift-dependent obscuration fraction is considered,
e.g. following~\cite{hasinger2008}, we find that the comparison
between our results and observed LFs remarkably improves. In both
energy bands, the simulated LFs decrease at low luminosities and are
in overall good agreement with observed LFs by~\cite{hasinger2005},
\cite{aird2010} and \cite{buchner2015} at almost all redshifts (see
Fig.~\ref{fig:sxr_hxr_lfs}).  However, we still note a discrepancy in
the SXR LFs at~$z=2.3$.

The importance of producing realistic X-ray catalogs of simulated AGN
in order to compare the intrinsic population of simulated SMBHs with
the statistical properties of observed AGN is also discussed in a
recent study by~\cite{OWLS2017} on the cosmo-OWLS simulations. They
also include a modelling of the obscuration fraction which depends on
redshift and luminosity of the AGN, showing that the AGN in the
cosmo-OWLS simulations reproduce very well the observational data. In
addition to LFs, \cite{OWLS2017} focus on the correlation function of
AGN in the mock XMM-Newton X-ray catalogs and on the comparison to
observed Eddington ratio distribution, finding as well a good match
with observations. Differently from their work, here we include
self-consistently the modelling of the AGN X-ray emission depending on
the intrinsic luminosity predicted by the simulations, and assuming an
observationally-based torus obscuration component for every AGN source
(including both Compton-thick and unobscured objects) in the simulation.

The use of mock X-ray observations is particularly important to study
the combined emission of AGN and ICM in clusters.  We dedicate the
second part of this analysis to explicitly studying the mock X-ray
emission of AGN in simulated clusters (Section~\ref{sec:results}).
This provides a prediction for the effects due to the presence of
AGN on the detection of clusters, as expected from eROSITA-like
observations.
Differently from the study of the absorbed LFs, where we do not
include any instrumental response in order to purely test our
modelling of the AGN emission including the torus obscuration
component, we generate complete eROSITA-like mock observations with
PHOX in order to investigate the observed AGN emission in galaxy
clusters.  Specifically, we employ the PHOX X-ray simulator to
generate eROSITA synthetic observations for a sample of galaxy
clusters with masses $\mfive > 3\times 10^{13}\msunh$, extracted from
the {\it Magenticum Pathfinder Simulation} at various redshifts
between $z=0$ and $z=2$. This allows us to explore and predict the
expected contamination from AGN emission to the ICM X-ray luminosity
of the hosting cluster, which will be very important in future X-ray
survey, especially of the high-redshift Universe, like eROSITA
(Section~\ref{sec:erosita}).
At low redshift ($z\lesssim 0.3$), we find that only for a small
fraction of clusters ($\sim 5\mbox{--}10\%$) the observed X-ray flux
from the AGN within the projected $\rfive$ radius is comparable to or
larger than the flux emitted by the whole ICM.
As expected, however, this fraction increases for increasing redshifts.
At redshifts $z\sim1\mbox{--}1.5$, the majority of our clusters are
faint and present a dominant AGN component in the X-ray
emission. Specifically, the flux observed from AGN and ICM is more
than a factor of 5 with respect to the flux from the ICM alone for
20--45\% of the sources.  If the observed SXR band is considered,
$\sim 34\%$ of our clusters show that the AGN+ICM flux is at least 10
times higher than the flux of the ICM.
  This result is consistent with the intrinsic prediction from the
  simulated catalogs, where the central AGN source(s) alone (residing
  within 5\% of the cluster virial radius) can emit an X-ray
  luminosity that is up 10 times higher than the $L_{X}$ of the whole ICM
  (see Fig.~\ref{fig:Lagn_Licm_sim}).
When only the subsample of massive clusters ($\mfive > 10^{14}\msunh$) is
considered, we find that the vast majority of them is not dominated by
the AGN emission, except for 15--25\% of the sources at
$z\sim1$--$1.2$.

The particular assumptions made in this analysis, for instance on the
observed distribution of column densities of the torus obscurer
component or on the luminosity- and redshift-dependent obscuration
fraction, can also moderately impact our conclusions.
We note, nevertheless, that updates to account for recent
observational improvements can be easily implemented into the PHOX
simulator, in future works.

\enlargethispage*{\baselineskip}

Next-generation of wide-area X-ray surveys, especially those dedicated
to exploring the high-$z$ Universe like eROSITA, will necessarily have
to deal with the ambiguity of detecting the diffuse emission from
galaxy clusters around powerful AGN sources, which might dominate the
X-ray observed flux. As we showed here, simulations allow to predict
the statistical occurrence of these cases.  Multi-wavelenght
observations can also play an important role in detecting the presence
of massive clusters around X-ray AGN sources, as discussed
by~\cite{green2017}.
Viceversa, spectroscopic followup of large number of X-ray sources
detected by eROSITA (such as those planned by SDSS-V,
\citealt{kollmeier2017}, or 4MOST, \citealt{dejong2014}) will allow
AGN to be identified with high reliability.
This will be crucial to investigate the role of
AGN and associated feedback within BCGs and clusters up to high redshifts,
where these phenomena are strictly connected to the thermodynamical
and chemical evolution of the cluster itself.

\appendix

\section{Mock eROSITA images}\label{app:mocks}
\begin{figure*}
  \centering
  \subfigure[$z=1$]{\includegraphics[width=0.47\textwidth]{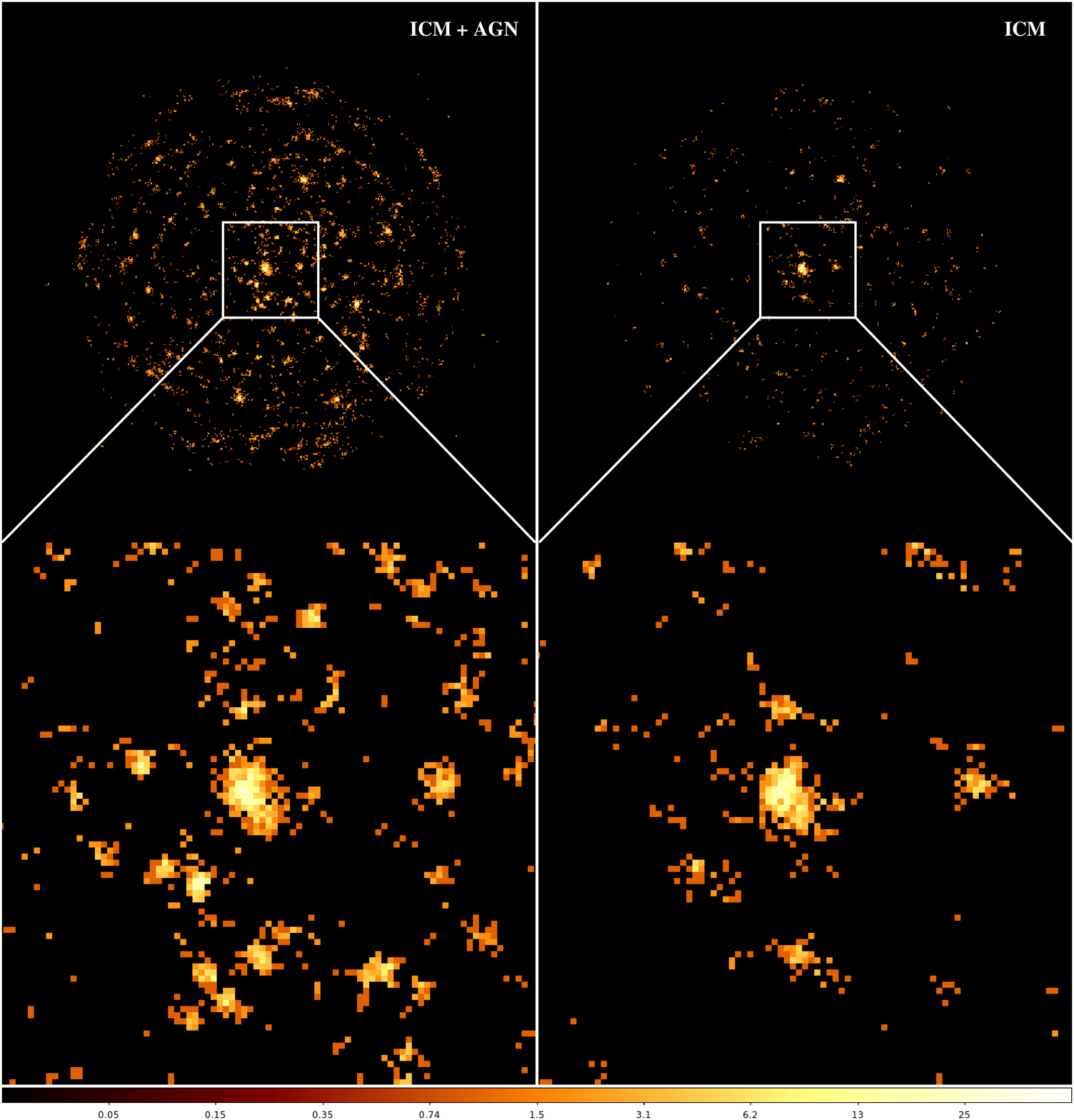}}\qquad
  \subfigure[$z=0.62$]{\includegraphics[width=0.47\textwidth]{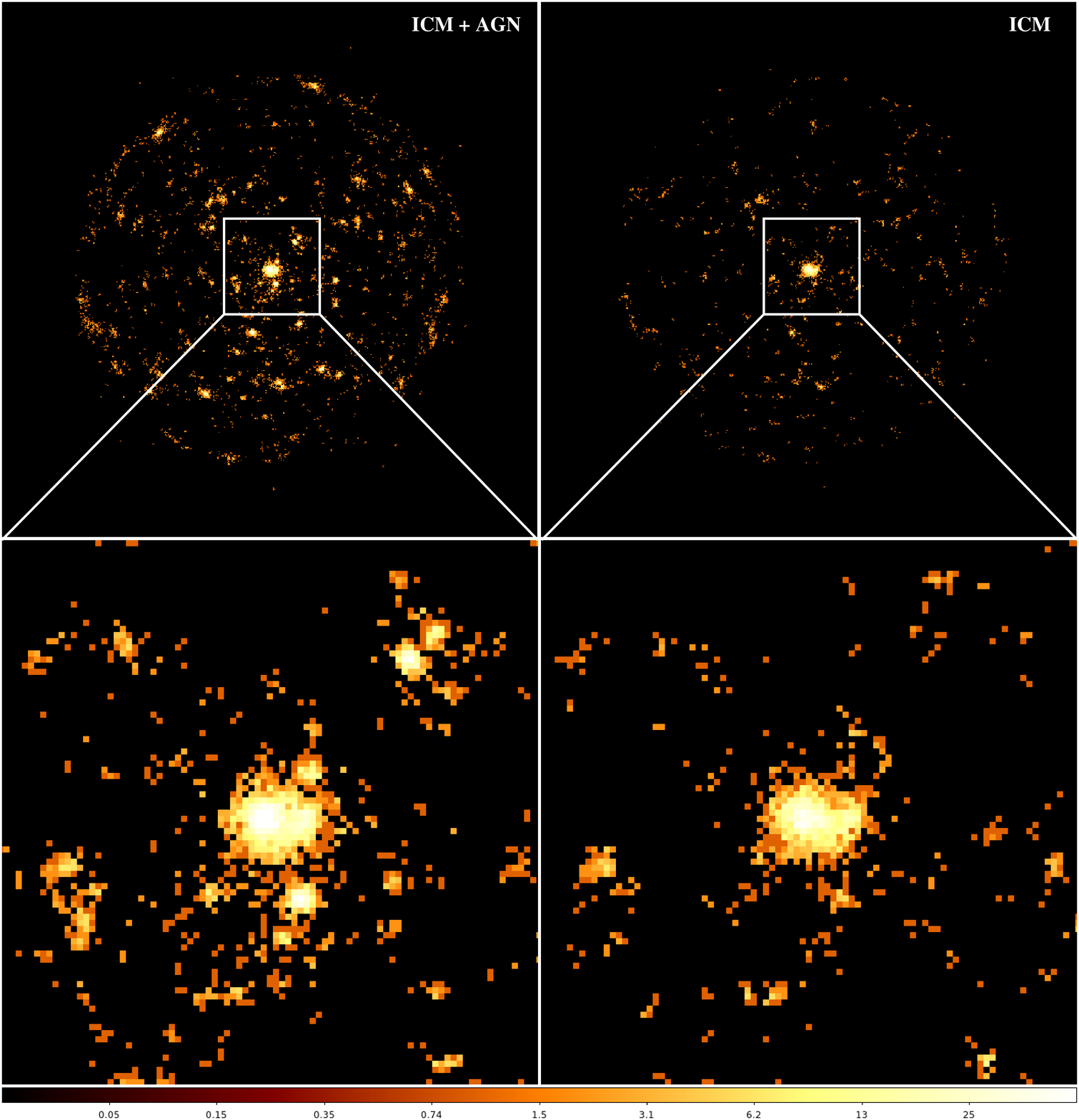}}
  \caption{eROSITA mock observations of two example clusters in our
    simulation box, one at $z=1$ (a) and one at $z=0.62$
    (b). In each figure we show the emission from AGN and
    ICM in the cluster (left insets) and the emission from the ICM
    only (right insets). In the bottom row insets we show the zoom
    onto the central region, as marked. The pointed observations were
    obtained with the standard SIXTE setup, without including the
    background and projecting along the line of sight for the whole
    simulation box lenght ($500$\,Mpc).\label{fig:mock}}
\end{figure*}

\begin{figure*}
  \centering
  \subfigure[$z=1$]{\includegraphics[width=0.47\textwidth]{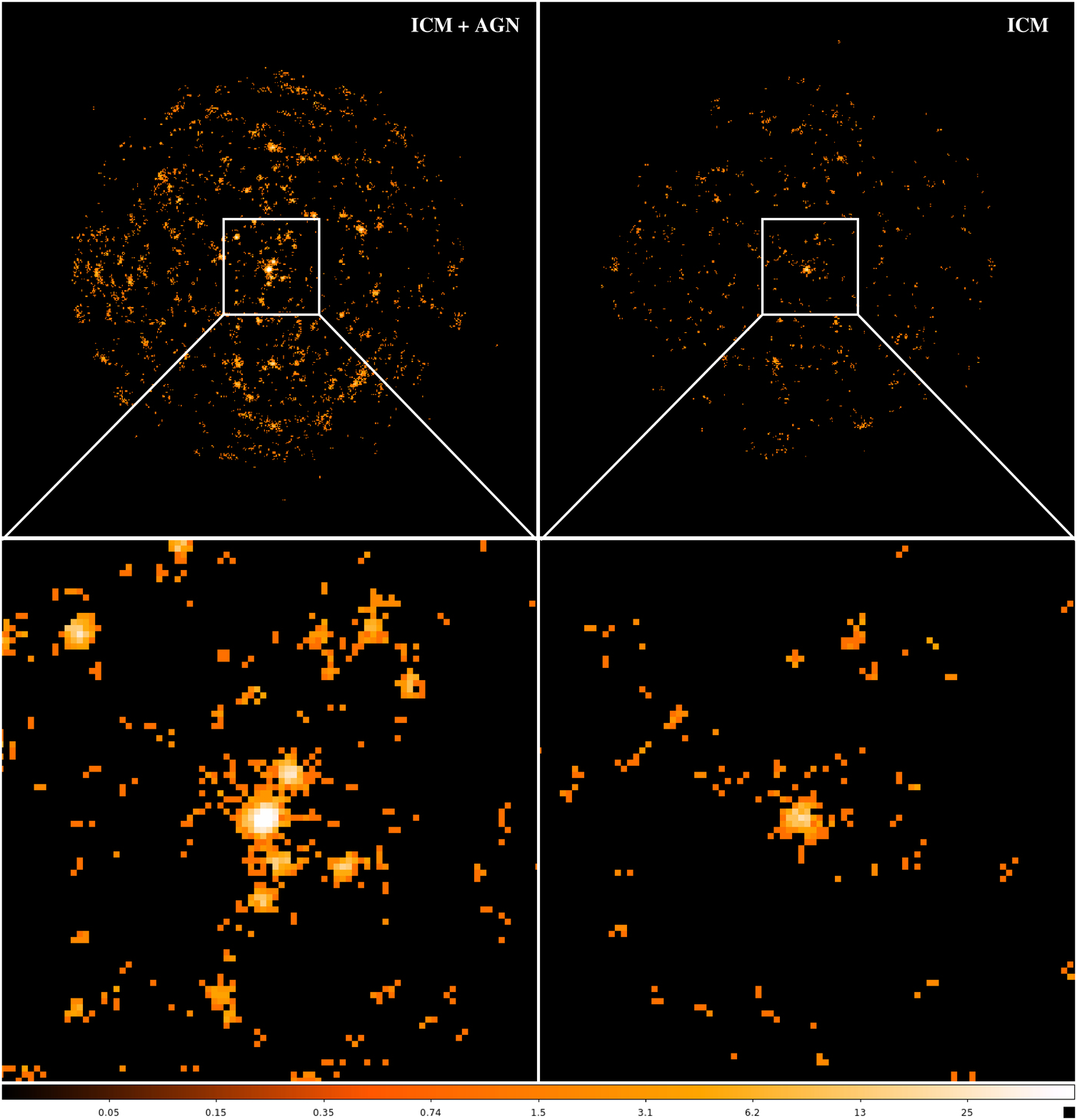}}\qquad
  \subfigure[$z=0.62$]{\includegraphics[width=0.47\textwidth]{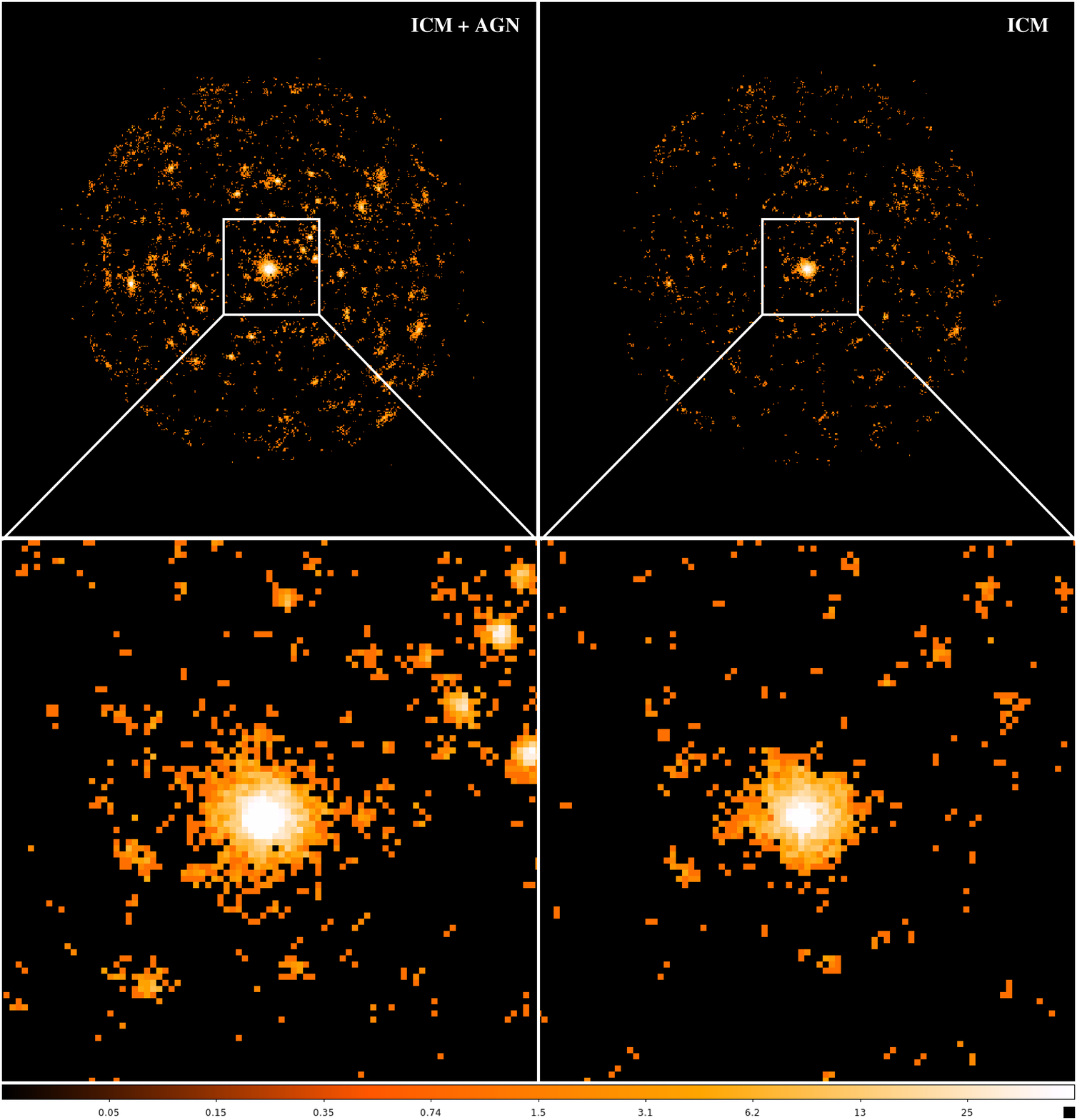}}
  \caption{Same as Fig.~\ref{fig:mock}, for two example clusters where
    the X-ray emission of the AGN is dominating over the ICM
    one.\label{fig:mock-agndom}}
\end{figure*}

As an example, in Fig.~\ref{fig:mock} we report the synthetic 10\,ks
eROSITA image of two example clusters in the sample, one at $z=1$
 and one at $z=0.62$.  In particular, we show the
images obtained from the combined emission of ICM and AGN sources in
each cluster (l.h.s.\ insets in each figure) and those for the
ICM only (r.h.s.\ insets).  These are two examples of detected
clusters, where the AGN emission is not dominant over the ICM.  In the
bottom-row panels we show the zoom onto the central region of the
pointing.
For comparison, we show in Fig.~\ref{fig:mock-agndom} two additional
example clusters where instead the AGN contribution to the X-ray
emission is dominant with respect to the ICM one. Specifically, from
the zoom onto the central region, one can clearly notice the increase
in the central X-ray surface brightness when the AGN emission is also
included (l.h.s.\ insets), with respect to the images obtained only
for the ICM (r.h.s.\ insets).

The images were obtained performing pointed observations with the
standard SIXTE setup, for which the PSF is rapidly degrading
(increasing) towards higher off-axis angles~\cite[see,
  e.g.][]{erosita-sciencebook}, as noticeable at the edges of the
pointings in the upper panels of Fig.~\ref{fig:mock}.  In reality,
however, observations like those investigated here will be rather
performed in survey (scanning) mode, which will results in an
effective average PSF Half-Energy Widthof $28"$ in the soft band and
$\sim 40"$ in the hard band.  In the current analysis these features
of the PSF do not play any major role, since we perform a pointed
observation for every single cluster, which typically resides in the
very central region of the FoV (see the zooms in
Figs.~\ref{fig:mock} and~\ref{fig:mock-agndom}).


\section*{Acknowledgments}
We are especially grateful for the support by M. Petkova through the
Computational Center for Particle and Astrophysics (C2PAP).
Computations have been performed at the `Leibniz-Rechenzentrum' with
CPU time assigned to the Project ``pr86re''.
The authors would like to thank the anonymous referee for
  constructive comments on the manuscript.
VB is thankful to Keith Arnaud for useful discussions and support on
the models embedded into \xspec, and to Johannes Buchner for kindly
providing the data reported in Fig.~3. VB also acknowledges Umberto
Maio and partial funding support from a grant of the German Research
Fundation (DFG), number 390015701.
KD acknowledges support by the DFG Transregio TR33 and the DFG
Cluster of Excellence ``Origin and Structure of the Universe.''


\bibliographystyle{mn2e}
\bibliography{bibl}

\begin{thebibliography}{}

\bibitem[\protect\citeauthoryear{{Aird}, {Coil}, {Georgakakis}, {Nandra},
  {Barro} \& {P{\'e}rez-Gonz{\'a}lez}}{{Aird} et~al.}{2015}]{aird2015}
{Aird} J.,  {Coil} A.~L.,  {Georgakakis} A.,  {Nandra} K.,  {Barro} G.,
  {P{\'e}rez-Gonz{\'a}lez} P.~G.,  2015, \mnras, 451, 1892

\bibitem[\protect\citeauthoryear{{Aird}, {Nandra}, {Laird}, {Georgakakis},
  {Ashby}, {Barmby}, {Coil}, {Huang}, {Koekemoer}, {Steidel} \&
  {Willmer}}{{Aird} et~al.}{2010}]{aird2010}
{Aird} J.,  {Nandra} K.,  {Laird} E.~S.,  {Georgakakis} A.,  {Ashby} M.~L.~N.,
  {Barmby} P.,  {Coil} A.~L.,  {Huang} J.-S.,  {Koekemoer} A.~M.,  {Steidel}
  C.~C.,    {Willmer} C.~N.~A.,  2010, \mnras, 401, 2531

\bibitem[\protect\citeauthoryear{{Allen}, {Evrard} \& {Mantz}}{{Allen}
  et~al.}{2011}]{allen2011}
{Allen} S.~W.,  {Evrard} A.~E.,    {Mantz} A.~B.,  2011, \araa, 49, 409

\bibitem[\protect\citeauthoryear{{Arnaud}}{{Arnaud}}{1996}]{xspec1996}
{Arnaud} K.~A.,  1996, in {G.~H.~Jacoby \& J.~Barnes} ed., Astronomical Data
  Analysis Software and Systems V Vol.~101 of Astronomical Society of the
  Pacific Conference Series, {XSPEC: The First Ten Years}.
pp 17--+

\bibitem[\protect\citeauthoryear{{Assef}, {Kochanek}, {Ashby}, {Brodwin},
  {Brown}, {Cool}, {Forman}, {Gonzalez}, {Hickox}, {Jannuzi}, {Jones}, {Le
  Floc'h}, {Moustakas}, {Murray} \& {Stern}}{{Assef} et~al.}{2011}]{assef2011}
{Assef} R.~J.,  {Kochanek} C.~S.,  {Ashby} M.~L.~N.,  {Brodwin} M.,  {Brown}
  M.~J.~I.,  {Cool} R.,  {Forman} W.,  {Gonzalez} A.~H.,  {Hickox} R.~C.,
  {Jannuzi} B.~T.,  {Jones} C.,  {Le Floc'h} E.,  {Moustakas} J.,  {Murray}
  S.~S.,    {Stern} D.,  2011, \apj, 728, 56

\bibitem[\protect\citeauthoryear{{Barai}, {Viel}, {Murante}, {Gaspari} \&
  {Borgani}}{{Barai} et~al.}{2014}]{barai2014}
{Barai} P.,  {Viel} M.,  {Murante} G.,  {Gaspari} M.,    {Borgani} S.,  2014,
  \mnras, 437, 1456

\bibitem[\protect\citeauthoryear{{Barger} \& {Cowie}}{{Barger} \&
  {Cowie}}{2005}]{barger2005}
{Barger} A.~J.,  {Cowie} L.~L.,  2005, \apj, 635, 115

\bibitem[\protect\citeauthoryear{{Beck}, {Murante}, {Arth}, {Remus}, {Teklu},
  {Donnert}, {Planelles}, {Beck}, {F{\"o}rster}, {Imgrund}, {Dolag} \&
  {Borgani}}{{Beck} et~al.}{2016}]{beck2016}
{Beck} A.~M.,  {Murante} G.,  {Arth} A.,  {Remus} R.-S.,  {Teklu} A.~F.,
  {Donnert} J.~M.~F.,  {Planelles} S.,  {Beck} M.~C.,  {F{\"o}rster} P.,
  {Imgrund} M.,  {Dolag} K.,    {Borgani} S.,  2016, \mnras, 455, 2110

\bibitem[\protect\citeauthoryear{{Benson}, {Bower}, {Frenk}, {Lacey}, {Baugh}
  \& {Cole}}{{Benson} et~al.}{2003}]{benson2003}
{Benson} A.~J.,  {Bower} R.~G.,  {Frenk} C.~S.,  {Lacey} C.~G.,  {Baugh} C.~M.,
     {Cole} S.,  2003, \apj, 599, 38

\bibitem[\protect\citeauthoryear{{Biffi}, {Dolag} \& {B{\"o}hringer}}{{Biffi}
  et~al.}{2013}]{biffi2013}
{Biffi} V.,  {Dolag} K.,    {B{\"o}hringer} H.,  2013, \mnras, 428, 1395

\bibitem[\protect\citeauthoryear{{Biffi}, {Dolag}, {B{\"o}hringer} \&
  {Lemson}}{{Biffi} et~al.}{2012}]{biffi2012}
{Biffi} V.,  {Dolag} K.,  {B{\"o}hringer} H.,    {Lemson} G.,  2012, \mnras,
  420, 3545

\bibitem[\protect\citeauthoryear{{Biffi}, {Planelles}, {Borgani}, {Fabjan},
  {Rasia}, {Murante}, {Tornatore}, {Dolag}, {Granato}, {Gaspari} \&
  {Beck}}{{Biffi} et~al.}{2017}]{biffi2017}
{Biffi} V.,  {Planelles} S.,  {Borgani} S.,  {Fabjan} D.,  {Rasia} E.,
  {Murante} G.,  {Tornatore} L.,  {Dolag} K.,  {Granato} G.~L.,  {Gaspari} M.,
    {Beck} A.~M.,  2017, \mnras, 468, 531

\bibitem[\protect\citeauthoryear{{Biffi}, {Sembolini}, {De Petris},
  {Valdarnini}, {Yepes} \& {Gottl{\"o}ber}}{{Biffi} et~al.}{2014}]{biffi2014}
{Biffi} V.,  {Sembolini} F.,  {De Petris} M.,  {Valdarnini} R.,  {Yepes} G.,
  {Gottl{\"o}ber} S.,  2014, \mnras, 439, 588

\bibitem[\protect\citeauthoryear{{Biffi} \& {Valdarnini}}{{Biffi} \&
  {Valdarnini}}{2015}]{biffi2015}
{Biffi} V.,  {Valdarnini} R.,  2015, \mnras, 446, 2802

\bibitem[\protect\citeauthoryear{{Bongiorno} \& {et al.}}{{Bongiorno} \& {et
  al.}}{2007}]{bongiorno2007}
{Bongiorno} A.,  {et al.} 2007, \aap, 472, 443

\bibitem[\protect\citeauthoryear{{Booth} \& {Schaye}}{{Booth} \&
  {Schaye}}{2009}]{booth2009}
{Booth} C.~M.,  {Schaye} J.,  2009, \mnras, 398, 53

\bibitem[\protect\citeauthoryear{{Booth} \& {Schaye}}{{Booth} \&
  {Schaye}}{2011}]{booth2011}
{Booth} C.~M.,  {Schaye} J.,  2011, \mnras, 413, 1158

\bibitem[\protect\citeauthoryear{{Boyle}, {Griffiths}, {Shanks}, {Stewart} \&
  {Georgantopoulos}}{{Boyle} et~al.}{1993}]{boyle1993}
{Boyle} B.~J.,  {Griffiths} R.~E.,  {Shanks} T.,  {Stewart} G.~C.,
  {Georgantopoulos} I.,  1993, \mnras, 260, 49

\bibitem[\protect\citeauthoryear{{Boyle}, {Shanks}, {Croom}, {Smith}, {Miller},
  {Loaring} \& {Heymans}}{{Boyle} et~al.}{2000}]{boyle2000}
{Boyle} B.~J.,  {Shanks} T.,  {Croom} S.~M.,  {Smith} R.~J.,  {Miller} L.,
  {Loaring} N.,    {Heymans} C.,  2000, \mnras, 317, 1014

\bibitem[\protect\citeauthoryear{{Boyle}, {Shanks}, {Georgantopoulos},
  {Stewart} \& {Griffiths}}{{Boyle} et~al.}{1994}]{boyle1994}
{Boyle} B.~J.,  {Shanks} T.,  {Georgantopoulos} I.,  {Stewart} G.~C.,
  {Griffiths} R.~E.,  1994, \mnras, 271, 639

\bibitem[\protect\citeauthoryear{{Buchner}, {Georgakakis}, {Nandra},
  {Brightman}, {Menzel}, {Liu}, {Hsu}, {Salvato}, {Rangel}, {Aird}, {Merloni}
  \& {Ross}}{{Buchner} et~al.}{2015}]{buchner2015}
{Buchner} J.,  {Georgakakis} A.,  {Nandra} K.,  {Brightman} M.,  {Menzel}
  M.-L.,  {Liu} Z.,  {Hsu} L.-T.,  {Salvato} M.,  {Rangel} C.,  {Aird} J.,
  {Merloni} A.,    {Ross} N.,  2015, \apj, 802, 89

\bibitem[\protect\citeauthoryear{{Buchner}, {Georgakakis}, {Nandra}, {Hsu},
  {Rangel}, {Brightman}, {Merloni}, {Salvato}, {Donley} \&
  {Kocevski}}{{Buchner} et~al.}{2014}]{buchner2014}
{Buchner} J.,  {Georgakakis} A.,  {Nandra} K.,  {Hsu} L.,  {Rangel} C.,
  {Brightman} M.,  {Merloni} A.,  {Salvato} M.,  {Donley} J.,    {Kocevski} D.,
   2014, \aap, 564, A125

\bibitem[\protect\citeauthoryear{{Churazov}, {Sazonov}, {Sunyaev}, {Forman},
  {Jones} \& {B{\"o}hringer}}{{Churazov} et~al.}{2005}]{churazov2005}
{Churazov} E.,  {Sazonov} S.,  {Sunyaev} R.,  {Forman} W.,  {Jones} C.,
  {B{\"o}hringer} H.,  2005, \mnras, 363, L91

\bibitem[\protect\citeauthoryear{{Croom}, {Richards}, {Shanks}, {Boyle},
  {Strauss}, {Myers}, {Nichol}, {Pimbblet}, {Ross}, {Schneider}, {Sharp} \&
  {Wake}}{{Croom} et~al.}{2009}]{croom2009}
{Croom} S.~M.,  {Richards} G.~T.,  {Shanks} T.,  {Boyle} B.~J.,  {Strauss}
  M.~A.,  {Myers} A.~D.,  {Nichol} R.~C.,  {Pimbblet} K.~A.,  {Ross} N.~P.,
  {Schneider} D.~P.,  {Sharp} R.~G.,    {Wake} D.~A.,  2009, \mnras, 399, 1755

\bibitem[\protect\citeauthoryear{{Croton}, {Springel}, {White}, {De Lucia},
  {Frenk}, {Gao}, {Jenkins}, {Kauffmann}, {Navarro} \& {Yoshida}}{{Croton}
  et~al.}{2006}]{croton2006}
{Croton} D.~J.,  {Springel} V.,  {White} S.~D.~M.,  {De Lucia} G.,  {Frenk}
  C.~S.,  {Gao} L.,  {Jenkins} A.,  {Kauffmann} G.,  {Navarro} J.~F.,
  {Yoshida} N.,  2006, \mnras, 365, 11

\bibitem[\protect\citeauthoryear{{Cui}, {Power}, {Biffi}, {Borgani}, {Murante},
  {Fabjan}, {Knebe}, {Lewis} \& {Poole}}{{Cui} et~al.}{2016}]{cui2016}
{Cui} W.,  {Power} C.,  {Biffi} V.,  {Borgani} S.,  {Murante} G.,  {Fabjan} D.,
   {Knebe} A.,  {Lewis} G.~F.,    {Poole} G.~B.,  2016, \mnras, 456, 2566

\bibitem[\protect\citeauthoryear{{David}, {Nulsen}, {McNamara}, {Forman},
  {Jones}, {Ponman}, {Robertson} \& {Wise}}{{David} et~al.}{2001}]{david2001}
{David} L.~P.,  {Nulsen} P.~E.~J.,  {McNamara} B.~R.,  {Forman} W.,  {Jones}
  C.,  {Ponman} T.,  {Robertson} B.,    {Wise} M.,  2001, \apj, 557, 546

\bibitem[\protect\citeauthoryear{{de Jong}, {Barden}, {Bellido-Tirado},
  {Brynnel}, {Chiappini}, {Depagne}, {Haynes}, {Johl}, {Phillips}, {Schnurr},
  {Schwope}, {Walcher} \& {et al.}}{{de Jong} et~al.}{2014}]{dejong2014}
{de Jong} R.~S.,  {Barden} S.,  {Bellido-Tirado} O.,  {Brynnel} J.,
  {Chiappini} C.,  {Depagne} {\'E}.,  {Haynes} R.,  {Johl} D.,  {Phillips}
  D.~P.,  {Schnurr} O.,  {Schwope} A.~D.,  {Walcher} J.,    {et al.} 2014, in
  Ground-based and Airborne Instrumentation for Astronomy V Vol.~9147 of
  \procspie, {4MOST: 4-metre Multi-Object Spectroscopic Telescope}.
p. 91470M

\bibitem[\protect\citeauthoryear{{Degraf}, {Di Matteo} \& {Springel}}{{Degraf}
  et~al.}{2010}]{degraf2010}
{Degraf} C.,  {Di Matteo} T.,    {Springel} V.,  2010, \mnras, 402, 1927

\bibitem[\protect\citeauthoryear{{Degraf}, {Di Matteo} \& {Springel}}{{Degraf}
  et~al.}{2011}]{degraf2011}
{Degraf} C.,  {Di Matteo} T.,    {Springel} V.,  2011, \mnras, 413, 1383

\bibitem[\protect\citeauthoryear{{Dehnen} \& {Aly}}{{Dehnen} \&
  {Aly}}{2012}]{dehnen2012}
{Dehnen} W.,  {Aly} H.,  2012, \mnras, 425, 1068

\bibitem[\protect\citeauthoryear{{Di Matteo}, {Colberg}, {Springel},
  {Hernquist} \& {Sijacki}}{{Di Matteo} et~al.}{2008}]{dimatteo2008}
{Di Matteo} T.,  {Colberg} J.,  {Springel} V.,  {Hernquist} L.,    {Sijacki}
  D.,  2008, \apj, 676, 33

\bibitem[\protect\citeauthoryear{{Di Matteo}, {Khandai}, {DeGraf}, {Feng},
  {Croft}, {Lopez} \& {Springel}}{{Di Matteo} et~al.}{2012}]{dimatteo2012}
{Di Matteo} T.,  {Khandai} N.,  {DeGraf} C.,  {Feng} Y.,  {Croft} R.~A.~C.,
  {Lopez} J.,    {Springel} V.,  2012, \apjl, 745, L29

\bibitem[\protect\citeauthoryear{{Di Matteo}, {Springel} \& {Hernquist}}{{Di
  Matteo} et~al.}{2005}]{dimatteo2005}
{Di Matteo} T.,  {Springel} V.,    {Hernquist} L.,  2005, \nat, 433, 604

\bibitem[\protect\citeauthoryear{{Dolag}, {Jubelgas}, {Springel}, {Borgani} \&
  {Rasia}}{{Dolag} et~al.}{2004}]{dolag2004}
{Dolag} K.,  {Jubelgas} M.,  {Springel} V.,  {Borgani} S.,    {Rasia} E.,
  2004, \apjl, 606, L97

\bibitem[\protect\citeauthoryear{{Dolag} \& {Stasyszyn}}{{Dolag} \&
  {Stasyszyn}}{2009}]{dolag_stasyszyn2009}
{Dolag} K.,  {Stasyszyn} F.,  2009, \mnras, 398, 1678

\bibitem[\protect\citeauthoryear{{Dolag}, {Vazza}, {Brunetti} \&
  {Tormen}}{{Dolag} et~al.}{2005}]{dolag2005}
{Dolag} K.,  {Vazza} F.,  {Brunetti} G.,    {Tormen} G.,  2005, \mnras, 364,
  753

\bibitem[\protect\citeauthoryear{{Fabian}}{{Fabian}}{1994}]{fabian1994}
{Fabian} A.~C.,  1994, \araa, 32, 277

\bibitem[\protect\citeauthoryear{{Fabian}}{{Fabian}}{2012}]{fabian2012}
{Fabian} A.~C.,  2012, \araa, 50, 455

\bibitem[\protect\citeauthoryear{{Fabian}, {Nulsen} \& {Canizares}}{{Fabian}
  et~al.}{1984}]{fabian1984}
{Fabian} A.~C.,  {Nulsen} P.~E.~J.,    {Canizares} C.~R.,  1984, \nat, 310, 733

\bibitem[\protect\citeauthoryear{{Fabian}, {Sanders}, {Allen}, {Crawford},
  {Iwasawa}, {Johnstone}, {Schmidt} \& {Taylor}}{{Fabian}
  et~al.}{2003}]{fabian2003}
{Fabian} A.~C.,  {Sanders} J.~S.,  {Allen} S.~W.,  {Crawford} C.~S.,  {Iwasawa}
  K.,  {Johnstone} R.~M.,  {Schmidt} R.~W.,    {Taylor} G.~B.,  2003, \mnras,
  344, L43

\bibitem[\protect\citeauthoryear{{Fabjan}, {Borgani}, {Tornatore}, {Saro},
  {Murante} \& {Dolag}}{{Fabjan} et~al.}{2010}]{fabjan2010}
{Fabjan} D.,  {Borgani} S.,  {Tornatore} L.,  {Saro} A.,  {Murante} G.,
  {Dolag} K.,  2010, \mnras, 401, 1670

\bibitem[\protect\citeauthoryear{{Ferrarese} \& {Ford}}{{Ferrarese} \&
  {Ford}}{2005}]{ferrarese2005}
{Ferrarese} L.,  {Ford} H.,  2005, \ssr, 116, 523

\bibitem[\protect\citeauthoryear{{Ferrarese} \& {Merritt}}{{Ferrarese} \&
  {Merritt}}{2000}]{ferrarese2000}
{Ferrarese} L.,  {Merritt} D.,  2000, \apjl, 539, L9

\bibitem[\protect\citeauthoryear{{Fiore}, {Puccetti}, {Grazian}, {Menci},
  {Shankar}, {Santini}, {Piconcelli}, {Koekemoer}, {Fontana}, {Boutsia},
  {Castellano}, {Lamastra}, {Malacaria}, {Feruglio}, {Mathur}, {Miller} \&
  {Pannella}}{{Fiore} et~al.}{2012}]{fiore2012}
{Fiore} F.,  {Puccetti} S.,  {Grazian} A.,  {Menci} N.,  {Shankar} F.,
  {Santini} P.,  {Piconcelli} E.,  {Koekemoer} A.~M.,  {Fontana} A.,  {Boutsia}
  K.,  {Castellano} M.,  {Lamastra} A.,  {Malacaria} C.,  {Feruglio} C.,
  {Mathur} S.,  {Miller} N.,    {Pannella} M.,  2012, \aap, 537, A16

\bibitem[\protect\citeauthoryear{{Fotopoulou}, {Buchner}, {Georgantopoulos},
  {Hasinger}, {Salvato}, {Georgakakis}, {Cappelluti}, {Ranalli}, {Hsu},
  {Brusa}, {Comastri}, {Miyaji}, {Nandra}, {Aird} \& {Paltani}}{{Fotopoulou}
  et~al.}{2016}]{fotopoulou2016}
{Fotopoulou} S.,  {Buchner} J.,  {Georgantopoulos} I.,  {Hasinger} G.,
  {Salvato} M.,  {Georgakakis} A.,  {Cappelluti} N.,  {Ranalli} P.,  {Hsu}
  L.~T.,  {Brusa} M.,  {Comastri} A.,  {Miyaji} T.,  {Nandra} K.,  {Aird} J.,
   {Paltani} S.,  2016, \aap, 587, A142

\bibitem[\protect\citeauthoryear{{Gebhardt}, {Bender}, {Bower}, {Dressler},
  {Faber}, {Filippenko}, {Green}, {Grillmair}, {Ho}, {Kormendy}, {Lauer},
  {Magorrian}, {Pinkney}, {Richstone} \& {Tremaine}}{{Gebhardt}
  et~al.}{2000}]{gebhardt2000}
{Gebhardt} K.,  {Bender} R.,  {Bower} G.,  {Dressler} A.,  {Faber} S.~M.,
  {Filippenko} A.~V.,  {Green} R.,  {Grillmair} C.,  {Ho} L.~C.,  {Kormendy}
  J.,  {Lauer} T.~R.,  {Magorrian} J.,  {Pinkney} J.,  {Richstone} D.,
  {Tremaine} S.,  2000, \apjl, 539, L13

\bibitem[\protect\citeauthoryear{{Gitti}, {Brighenti} \& {McNamara}}{{Gitti}
  et~al.}{2012}]{gitti2012}
{Gitti} M.,  {Brighenti} F.,    {McNamara} B.~R.,  2012, Advances in Astronomy,
  2012, 950641

\bibitem[\protect\citeauthoryear{{Green}, {Edge}, {Ebeling}, {Burgett},
  {Draper}, {Kaiser}, {Kudritzki}, {Magnier}, {Metcalfe}, {Wainscoat} \&
  {Waters}}{{Green} et~al.}{2017}]{green2017}
{Green} T.~S.,  {Edge} A.~C.,  {Ebeling} H.,  {Burgett} W.~S.,  {Draper} P.~W.,
   {Kaiser} N.,  {Kudritzki} R.-P.,  {Magnier} E.~A.,  {Metcalfe} N.,
  {Wainscoat} R.~J.,    {Waters} C.,  2017, \mnras, 465, 4872

\bibitem[\protect\citeauthoryear{{Haardt} \& {Madau}}{{Haardt} \&
  {Madau}}{2001}]{haardt2001}
{Haardt} F.,  {Madau} P.,  2001, in {Neumann} D.~M.,  {Tran} J.~T.~V.,  eds,
  Clusters of Galaxies and the High Redshift Universe Observed in X-rays
  {Modelling the UV/X-ray cosmic background with CUBA}

\bibitem[\protect\citeauthoryear{{Haas}, {Schaye}, {Booth}, {Dalla Vecchia},
  {Springel}, {Theuns} \& {Wiersma}}{{Haas} et~al.}{2013}]{haas2013}
{Haas} M.~R.,  {Schaye} J.,  {Booth} C.~M.,  {Dalla Vecchia} C.,  {Springel}
  V.,  {Theuns} T.,    {Wiersma} R.~P.~C.,  2013, \mnras, 435, 2931

\bibitem[\protect\citeauthoryear{{Hahn}, {Martizzi}, {Wu}, {Evrard}, {Teyssier}
  \& {Wechsler}}{{Hahn} et~al.}{2015}]{hahn2015}
{Hahn} O.,  {Martizzi} D.,  {Wu} H.-Y.,  {Evrard} A.~E.,  {Teyssier} R.,
  {Wechsler} R.~H.,  2015, ArXiv e-prints

\bibitem[\protect\citeauthoryear{{Hasinger}}{{Hasinger}}{2008}]{hasinger2008}
{Hasinger} G.,  2008, \aap, 490, 905

\bibitem[\protect\citeauthoryear{{Hasinger}, {Miyaji} \& {Schmidt}}{{Hasinger}
  et~al.}{2005}]{hasinger2005}
{Hasinger} G.,  {Miyaji} T.,    {Schmidt} M.,  2005, \aap, 441, 417

\bibitem[\protect\citeauthoryear{{Hirschmann}, {Dolag}, {Saro}, {Bachmann},
  {Borgani} \& {Burkert}}{{Hirschmann} et~al.}{2014}]{H14}
{Hirschmann} M.,  {Dolag} K.,  {Saro} A.,  {Bachmann} L.,  {Borgani} S.,
  {Burkert} A.,  2014, \mnras, 442, 2304

\bibitem[\protect\citeauthoryear{{Hopkins}, {Hernquist}, {Cox}, {Robertson},
  {Di Matteo} \& {Springel}}{{Hopkins} et~al.}{2006}]{hopkins2006}
{Hopkins} P.~F.,  {Hernquist} L.,  {Cox} T.~J.,  {Robertson} B.,  {Di Matteo}
  T.,    {Springel} V.,  2006, \apj, 639, 700

\bibitem[\protect\citeauthoryear{{Hopkins}, {Richards} \&
  {Hernquist}}{{Hopkins} et~al.}{2007}]{hopkins07}
{Hopkins} P.~F.,  {Richards} G.~T.,    {Hernquist} L.,  2007, \apj, 654, 731

\bibitem[\protect\citeauthoryear{{Kollmeier}, {Zasowski}, {Rix}, {Johns},
  {Anderson}, {Drory}, {Johnson}, {Pogge}, {Bird}, {Blanc}, {Brownstein},
  {Crane}, {De Lee}, {Klaene}, {Kreckel}, {MacDonald}, {Merloni} \& {et
  al.}}{{Kollmeier} et~al.}{2017}]{kollmeier2017}
{Kollmeier} J.~A.,  {Zasowski} G.,  {Rix} H.-W.,  {Johns} M.,  {Anderson}
  S.~F.,  {Drory} N.,  {Johnson} J.~A.,  {Pogge} R.~W.,  {Bird} J.~C.,  {Blanc}
  G.~A.,  {Brownstein} J.~R.,  {Crane} J.~D.,  {De Lee} N.~M.,  {Klaene} M.~A.,
   {Kreckel} K.,  {MacDonald} N.,  {Merloni} A.,    {et al.} 2017, ArXiv
  e-prints

\bibitem[\protect\citeauthoryear{{Kormendy} \& {Ho}}{{Kormendy} \&
  {Ho}}{2013}]{kormendy2013}
{Kormendy} J.,  {Ho} L.~C.,  2013, \araa, 51, 511

\bibitem[\protect\citeauthoryear{{Koulouridis}, {Faccioli}, {Le Brun},
  {Plionis}, {McCarthy}, {Pierre}, {Akylas}, {Georgantopoulos}, {Paltani},
  {Lidman}, {Fotopoulou}, {Vignali}, {Pacaud} \& {Ranalli}}{{Koulouridis}
  et~al.}{2017}]{OWLS2017}
{Koulouridis} E.,  {Faccioli} L.,  {Le Brun} A.~M.~C.,  {Plionis} M.,
  {McCarthy} I.~G.,  {Pierre} M.,  {Akylas} A.,  {Georgantopoulos} I.,
  {Paltani} S.,  {Lidman} C.,  {Fotopoulou} S.,  {Vignali} C.,  {Pacaud} F.,
  {Ranalli} P.,  2017, ArXiv e-prints

\bibitem[\protect\citeauthoryear{{La Franca}, {Fiore}, {Comastri}, {Perola},
  {Sacchi}, {Brusa}, {Cocchia}, {Feruglio}, {Matt}, {Vignali}, {Carangelo},
  {Ciliegi}, {Lamastra}, {Maiolino}, {Mignoli}, {Molendi} \& {Puccetti}}{{La
  Franca} et~al.}{2005}]{lafranca2005}
{La Franca} F.,  {Fiore} F.,  {Comastri} A.,  {Perola} G.~C.,  {Sacchi} N.,
  {Brusa} M.,  {Cocchia} F.,  {Feruglio} C.,  {Matt} G.,  {Vignali} C.,
  {Carangelo} N.,  {Ciliegi} P.,  {Lamastra} A.,  {Maiolino} R.,  {Mignoli} M.,
   {Molendi} S.,    {Puccetti} S.,  2005, \apj, 635, 864

\bibitem[\protect\citeauthoryear{{Liu}, {Tozzi}, {Wang}, {Brandt}, {Vignali},
  {Xue}, {Schneider}, {Comastri} \& {et al.}}{{Liu} et~al.}{2017}]{liu2017}
{Liu} T.,  {Tozzi} P.,  {Wang} J.-X.,  {Brandt} W.~N.,  {Vignali} C.,  {Xue}
  Y.,  {Schneider} D.~P.,  {Comastri} A.,    {et al.} 2017, \apjs, 232, 8

\bibitem[\protect\citeauthoryear{{Maccacaro}, {della Ceca}, {Gioia}, {Morris},
  {Stocke} \& {Wolter}}{{Maccacaro} et~al.}{1991}]{maccacaro1991}
{Maccacaro} T.,  {della Ceca} R.,  {Gioia} I.~M.,  {Morris} S.~L.,  {Stocke}
  J.~T.,    {Wolter} A.,  1991, \apj, 374, 117

\bibitem[\protect\citeauthoryear{{Maccacaro}, {Gioia}, {Avni}, {Giommi},
  {Griffiths}, {Liebert}, {Stocke} \& {Danziger}}{{Maccacaro}
  et~al.}{1983}]{maccacaro1983}
{Maccacaro} T.,  {Gioia} I.~M.,  {Avni} Y.,  {Giommi} P.,  {Griffiths} R.~E.,
  {Liebert} J.,  {Stocke} J.,    {Danziger} J.,  1983, \apjl, 266, L73

\bibitem[\protect\citeauthoryear{{Maccacaro}, {Gioia} \& {Stocke}}{{Maccacaro}
  et~al.}{1984}]{maccacaro1984}
{Maccacaro} T.,  {Gioia} I.~M.,    {Stocke} J.~T.,  1984, \apj, 283, 486

\bibitem[\protect\citeauthoryear{{Magorrian}, {Tremaine}, {Richstone},
  {Bender}, {Bower}, {Dressler}, {Faber}, {Gebhardt}, {Green}, {Grillmair},
  {Kormendy} \& {Lauer}}{{Magorrian} et~al.}{1998}]{magorrian1998}
{Magorrian} J.,  {Tremaine} S.,  {Richstone} D.,  {Bender} R.,  {Bower} G.,
  {Dressler} A.,  {Faber} S.~M.,  {Gebhardt} K.,  {Green} R.,  {Grillmair} C.,
  {Kormendy} J.,    {Lauer} T.,  1998, \aj, 115, 2285

\bibitem[\protect\citeauthoryear{{Maio}, {Dotti}, {Petkova}, {Perego} \&
  {Volonteri}}{{Maio} et~al.}{2013}]{maio2013}
{Maio} U.,  {Dotti} M.,  {Petkova} M.,  {Perego} A.,    {Volonteri} M.,  2013,
  \apj, 767, 37

\bibitem[\protect\citeauthoryear{{Marconi}, {Risaliti}, {Gilli}, {Hunt},
  {Maiolino} \& {Salvati}}{{Marconi} et~al.}{2004}]{marconi2004}
{Marconi} A.,  {Risaliti} G.,  {Gilli} R.,  {Hunt} L.~K.,  {Maiolino} R.,
  {Salvati} M.,  2004, \mnras, 351, 169

\bibitem[\protect\citeauthoryear{{Martizzi}, {Hahn}, {Wu}, {Evrard}, {Teyssier}
  \& {Wechsler}}{{Martizzi} et~al.}{2016}]{martizzi2016}
{Martizzi} D.,  {Hahn} O.,  {Wu} H.-Y.,  {Evrard} A.~E.,  {Teyssier} R.,
  {Wechsler} R.~H.,  2016, \mnras, 459, 4408

\bibitem[\protect\citeauthoryear{{McCarthy}, {Schaye}, {Bower}, {Ponman},
  {Booth}, {Dalla Vecchia} \& {Springel}}{{McCarthy}
  et~al.}{2011}]{mccarthy2011}
{McCarthy} I.~G.,  {Schaye} J.,  {Bower} R.~G.,  {Ponman} T.~J.,  {Booth}
  C.~M.,  {Dalla Vecchia} C.,    {Springel} V.,  2011, \mnras, 412, 1965

\bibitem[\protect\citeauthoryear{{McCarthy}, {Schaye}, {Ponman}, {Bower},
  {Booth}, {Dalla Vecchia}, {Crain}, {Springel}, {Theuns} \&
  {Wiersma}}{{McCarthy} et~al.}{2010}]{mccarthy2010}
{McCarthy} I.~G.,  {Schaye} J.,  {Ponman} T.~J.,  {Bower} R.~G.,  {Booth}
  C.~M.,  {Dalla Vecchia} C.,  {Crain} R.~A.,  {Springel} V.,  {Theuns} T.,
  {Wiersma} R.~P.~C.,  2010, \mnras, 406, 822

\bibitem[\protect\citeauthoryear{{McConnell} \& {Ma}}{{McConnell} \&
  {Ma}}{2013}]{mcconnell2013}
{McConnell} N.~J.,  {Ma} C.-P.,  2013, \apj, 764, 184

\bibitem[\protect\citeauthoryear{{McNamara}, {Nulsen}, {Wise}, {Rafferty},
  {Carilli}, {Sarazin} \& {Blanton}}{{McNamara} et~al.}{2005}]{mcnamara2005}
{McNamara} B.~R.,  {Nulsen} P.~E.~J.,  {Wise} M.~W.,  {Rafferty} D.~A.,
  {Carilli} C.,  {Sarazin} C.~L.,    {Blanton} E.~L.,  2005, \nat, 433, 45

\bibitem[\protect\citeauthoryear{{McNamara}, {Wise}, {Nulsen}, {David},
  {Sarazin}, {Bautz}, {Markevitch}, {Vikhlinin}, {Forman}, {Jones} \&
  {Harris}}{{McNamara} et~al.}{2000}]{mcnamara2000}
{McNamara} B.~R.,  {Wise} M.,  {Nulsen} P.~E.~J.,  {David} L.~P.,  {Sarazin}
  C.~L.,  {Bautz} M.,  {Markevitch} M.,  {Vikhlinin} A.,  {Forman} W.~R.,
  {Jones} C.,    {Harris} D.~E.,  2000, \apjl, 534, L135

\bibitem[\protect\citeauthoryear{{Merloni} \& {et al.}}{{Merloni} \& {et
  al.}}{2014}]{merloni2014}
{Merloni} A.,  {et al.} 2014, \mnras, 437, 3550

\bibitem[\protect\citeauthoryear{{Merloni} \& {Heinz}}{{Merloni} \&
  {Heinz}}{2008}]{merloni2008}
{Merloni} A.,  {Heinz} S.,  2008, \mnras, 388, 1011

\bibitem[\protect\citeauthoryear{{Merloni}, {Predehl}, {Becker},
  {B{\"o}hringer}, {Boller}, {Brunner}, {Brusa} \& {German eROSITA
  Consortium}}{{Merloni} et~al.}{2012}]{erosita-sciencebook}
{Merloni} A.,  {Predehl} P.,  {Becker} W.,  {B{\"o}hringer} H.,  {Boller} T.,
  {Brunner} H.,  {Brusa} M.,    {German eROSITA Consortium} t.,  2012, ArXiv
  e-prints

\bibitem[\protect\citeauthoryear{{Page}, {Carrera}, {Hasinger}, {Mason},
  {McMahon}, {Mittaz}, {Barcons}, {Carballo}, {Gonzalez-Serrano} \&
  {Perez-Fournon}}{{Page} et~al.}{1996}]{page1996}
{Page} M.~J.,  {Carrera} F.~J.,  {Hasinger} G.,  {Mason} K.~O.,  {McMahon}
  R.~G.,  {Mittaz} J.~P.~D.,  {Barcons} X.,  {Carballo} R.,  {Gonzalez-Serrano}
  I.,    {Perez-Fournon} I.,  1996, \mnras, 281, 579

\bibitem[\protect\citeauthoryear{{Peterson} \& {Fabian}}{{Peterson} \&
  {Fabian}}{2006}]{peterson2006}
{Peterson} J.~R.,  {Fabian} A.~C.,  2006, \physrep, 427, 1

\bibitem[\protect\citeauthoryear{{Peterson}, {Paerels}, {Kaastra}, {Arnaud},
  {Reiprich}, {Fabian}, {Mushotzky}, {Jernigan} \& {Sakelliou}}{{Peterson}
  et~al.}{2001}]{peterson2001}
{Peterson} J.~R.,  {Paerels} F.~B.~S.,  {Kaastra} J.~S.,  {Arnaud} M.,
  {Reiprich} T.~H.,  {Fabian} A.~C.,  {Mushotzky} R.~F.,  {Jernigan} J.~G.,
  {Sakelliou} I.,  2001, \aap, 365, L104

\bibitem[\protect\citeauthoryear{{Pillepich}, {Porciani} \&
  {Reiprich}}{{Pillepich} et~al.}{2012}]{pillepich2012}
{Pillepich} A.,  {Porciani} C.,    {Reiprich} T.~H.,  2012, \mnras, 422, 44

\bibitem[\protect\citeauthoryear{{Planelles}, {Fabjan}, {Borgani}, {Murante},
  {Rasia}, {Biffi}, {Truong}, {Ragone-Figueroa}, {Granato}, {Dolag},
  {Pierpaoli}, {Beck}, {Steinborn} \& {Gaspari}}{{Planelles}
  et~al.}{2017}]{planelles2017}
{Planelles} S.,  {Fabjan} D.,  {Borgani} S.,  {Murante} G.,  {Rasia} E.,
  {Biffi} V.,  {Truong} N.,  {Ragone-Figueroa} C.,  {Granato} G.~L.,  {Dolag}
  K.,  {Pierpaoli} E.,  {Beck} A.~M.,  {Steinborn} L.~K.,    {Gaspari} M.,
  2017, \mnras, 467, 3827

\bibitem[\protect\citeauthoryear{{Ragagnin}, {Dolag}, {Biffi}, {Cadolle Bel},
  {Hammer}, {Krukau}, {Petkova} \& {Steinborn}}{{Ragagnin}
  et~al.}{2017}]{ragagnin2017}
{Ragagnin} A.,  {Dolag} K.,  {Biffi} V.,  {Cadolle Bel} M.,  {Hammer} N.~J.,
  {Krukau} A.,  {Petkova} M.,    {Steinborn} D.,  2017, Astronomy and
  Computing, 20, 52

\bibitem[\protect\citeauthoryear{{Ranalli} \& {et al.}}{{Ranalli} \& {et
  al.}}{2016}]{ranalli2016}
{Ranalli} P.,  {et al.} 2016, \aap, 590, A80

\bibitem[\protect\citeauthoryear{{Rasia}, {Borgani}, {Murante}, {Planelles},
  {Beck}, {Biffi}, {Ragone-Figueroa}, {Granato}, {Steinborn} \&
  {Dolag}}{{Rasia} et~al.}{2015}]{rasia2015}
{Rasia} E.,  {Borgani} S.,  {Murante} G.,  {Planelles} S.,  {Beck} A.~M.,
  {Biffi} V.,  {Ragone-Figueroa} C.,  {Granato} G.~L.,  {Steinborn} L.~K.,
  {Dolag} K.,  2015, \apjl, 813, L17

\bibitem[\protect\citeauthoryear{{Richards} \& {et al.}}{{Richards} \& {et
  al.}}{2006}]{richards2006}
{Richards} G.~T.,  {et al.} 2006, \aj, 131, 2766

\bibitem[\protect\citeauthoryear{{Sarazin}}{{Sarazin}}{1986}]{sarazin1986}
{Sarazin} C.~L.,  1986, Reviews of Modern Physics, 58, 1

\bibitem[\protect\citeauthoryear{{Sijacki}, {Springel}, {Di Matteo} \&
  {Hernquist}}{{Sijacki} et~al.}{2007}]{sijacki2007}
{Sijacki} D.,  {Springel} V.,  {Di Matteo} T.,    {Hernquist} L.,  2007,
  \mnras, 380, 877

\bibitem[\protect\citeauthoryear{{Sijacki}, {Vogelsberger}, {Genel},
  {Springel}, {Torrey}, {Snyder}, {Nelson} \& {Hernquist}}{{Sijacki}
  et~al.}{2015}]{sijacki2015}
{Sijacki} D.,  {Vogelsberger} M.,  {Genel} S.,  {Springel} V.,  {Torrey} P.,
  {Snyder} G.~F.,  {Nelson} D.,    {Hernquist} L.,  2015, \mnras, 452, 575

\bibitem[\protect\citeauthoryear{{Silverman}, {Green}, {Barkhouse}, {Kim},
  {Kim}, {Wilkes}, {Cameron}, {Hasinger}, {Jannuzi}, {Smith}, {Smith} \&
  {Tananbaum}}{{Silverman} et~al.}{2008}]{silverman2008}
{Silverman} J.~D.,  {Green} P.~J.,  {Barkhouse} W.~A.,  {Kim} D.-W.,  {Kim} M.,
   {Wilkes} B.~J.,  {Cameron} R.~A.,  {Hasinger} G.,  {Jannuzi} B.~T.,  {Smith}
  M.~G.,  {Smith} P.~S.,    {Tananbaum} H.,  2008, \apj, 679, 118

\bibitem[\protect\citeauthoryear{{Simpson}}{{Simpson}}{2005}]{simpson2005}
{Simpson} C.,  2005, \mnras, 360, 565

\bibitem[\protect\citeauthoryear{{Smith}, {Brickhouse}, {Liedahl} \&
  {Raymond}}{{Smith} et~al.}{2001}]{apec2001}
{Smith} R.~K.,  {Brickhouse} N.~S.,  {Liedahl} D.~A.,    {Raymond} J.~C.,
  2001, \apjl, 556, L91

\bibitem[\protect\citeauthoryear{{Springel}}{{Springel}}{2005}]{springel2005}
{Springel} V.,  2005, \mnras, 364, 1105

\bibitem[\protect\citeauthoryear{{Springel}, {Di Matteo} \&
  {Hernquist}}{{Springel} et~al.}{2005}]{springeldimatteo2005}
{Springel} V.,  {Di Matteo} T.,    {Hernquist} L.,  2005, \mnras, 361, 776

\bibitem[\protect\citeauthoryear{{Springel} \& {Hernquist}}{{Springel} \&
  {Hernquist}}{2002}]{springel2002}
{Springel} V.,  {Hernquist} L.,  2002, \mnras, 333, 649

\bibitem[\protect\citeauthoryear{{Springel} \& {Hernquist}}{{Springel} \&
  {Hernquist}}{2003}]{springel2003}
{Springel} V.,  {Hernquist} L.,  2003, \mnras, 339, 289

\bibitem[\protect\citeauthoryear{{Steinborn}, {Dolag}, {Comerford},
  {Hirschmann}, {Remus} \& {Teklu}}{{Steinborn} et~al.}{2016}]{lisa2016}
{Steinborn} L.~K.,  {Dolag} K.,  {Comerford} J.~M.,  {Hirschmann} M.,  {Remus}
  R.-S.,    {Teklu} A.~F.,  2016, \mnras, 458, 1013

\bibitem[\protect\citeauthoryear{{Steinborn}, {Dolag}, {Hirschmann}, {Prieto}
  \& {Remus}}{{Steinborn} et~al.}{2015}]{lisa2015}
{Steinborn} L.~K.,  {Dolag} K.,  {Hirschmann} M.,  {Prieto} M.~A.,    {Remus}
  R.-S.,  2015, \mnras, 448, 1504

\bibitem[\protect\citeauthoryear{{Tornatore}, {Borgani}, {Dolag} \&
  {Matteucci}}{{Tornatore} et~al.}{2007}]{tornatore2007}
{Tornatore} L.,  {Borgani} S.,  {Dolag} K.,    {Matteucci} F.,  2007, \mnras,
  382, 1050

\bibitem[\protect\citeauthoryear{{Tornatore}, {Borgani}, {Springel},
  {Matteucci}, {Menci} \& {Murante}}{{Tornatore} et~al.}{2003}]{tornatore2003}
{Tornatore} L.,  {Borgani} S.,  {Springel} V.,  {Matteucci} F.,  {Menci} N.,
  {Murante} G.,  2003, \mnras, 342, 1025

\bibitem[\protect\citeauthoryear{{Tremaine}, {Gebhardt}, {Bender}, {Bower},
  {Dressler}, {Faber}, {Filippenko}, {Green}, {Grillmair}, {Ho}, {Kormendy},
  {Lauer}, {Magorrian}, {Pinkney} \& {Richstone}}{{Tremaine}
  et~al.}{2002}]{tremaine2002}
{Tremaine} S.,  {Gebhardt} K.,  {Bender} R.,  {Bower} G.,  {Dressler} A.,
  {Faber} S.~M.,  {Filippenko} A.~V.,  {Green} R.,  {Grillmair} C.,  {Ho}
  L.~C.,  {Kormendy} J.,  {Lauer} T.~R.,  {Magorrian} J.,  {Pinkney} J.,
  {Richstone} D.,  2002, \apj, 574, 740

\bibitem[\protect\citeauthoryear{{Ueda}, {Akiyama}, {Hasinger}, {Miyaji} \&
  {Watson}}{{Ueda} et~al.}{2014}]{ueda2014}
{Ueda} Y.,  {Akiyama} M.,  {Hasinger} G.,  {Miyaji} T.,    {Watson} M.~G.,
  2014, \apj, 786, 104

\bibitem[\protect\citeauthoryear{{Ueda}, {Akiyama}, {Ohta} \& {Miyaji}}{{Ueda}
  et~al.}{2003}]{ueda2003}
{Ueda} Y.,  {Akiyama} M.,  {Ohta} K.,    {Miyaji} T.,  2003, \apj, 598, 886

\bibitem[\protect\citeauthoryear{{Vogelsberger}, {Marinacci}, {Torrey},
  {Genel}, {Springel}, {Weinberger}, {Pakmor}, {Hernquist}, {Naiman},
  {Pillepich} \& {Nelson}}{{Vogelsberger} et~al.}{2017}]{vogelsberger2017}
{Vogelsberger} M.,  {Marinacci} F.,  {Torrey} P.,  {Genel} S.,  {Springel} V.,
  {Weinberger} R.,  {Pakmor} R.,  {Hernquist} L.,  {Naiman} J.,  {Pillepich}
  A.,    {Nelson} D.,  2017, ArXiv e-prints

\bibitem[\protect\citeauthoryear{{Voit} \& {Donahue}}{{Voit} \&
  {Donahue}}{2005}]{voit2005}
{Voit} G.~M.,  {Donahue} M.,  2005, \apj, 634, 955

\bibitem[\protect\citeauthoryear{{Wiersma}, {Schaye} \& {Smith}}{{Wiersma}
  et~al.}{2009}]{wiersma2009}
{Wiersma} R.~P.~C.,  {Schaye} J.,    {Smith} B.~D.,  2009, \mnras, 393, 99

\bibitem[\protect\citeauthoryear{{Wolf}, {Meisenheimer}, {Rix}, {Borch}, {Dye}
  \& {Kleinheinrich}}{{Wolf} et~al.}{2003}]{wolf2003}
{Wolf} C.,  {Meisenheimer} K.,  {Rix} H.-W.,  {Borch} A.,  {Dye} S.,
  {Kleinheinrich} M.,  2003, \aap, 401, 73

\bibitem[\protect\citeauthoryear{{Zdziarski}, {Poutanen} \&
  {Johnson}}{{Zdziarski} et~al.}{2000}]{Zdziarski2000}
{Zdziarski} A.~A.,  {Poutanen} J.,    {Johnson} W.~N.,  2000, \apj, 542, 703

\end{thebibliography}
%



\label{lastpage}
\end{document}